\documentclass[11pt]{article}
\usepackage{amsmath,amssymb,color,epsfig,cite}
\usepackage{xcolor}
\usepackage{mathrsfs}

\usepackage{mathtools}

\allowdisplaybreaks

\usepackage{amsfonts}

\newcommand{\hoch}[1]{$\, ^{#1}$}

\makeatletter
\@addtoreset{equation}{section}
\makeatother

\newcommand{\be}{\begin{equation}}
\newcommand{\ee}{\end{equation}}
\newcommand{\bea} {\begin{eqnarray}}
\newcommand{\eea}{\end{eqnarray}}
\newcommand{\nn}{\nonumber}

\def\ft#1#2{{\textstyle{\frac{\scriptstyle #1}{\scriptstyle #2} } }}
\def\fft#1#2{{\frac{#1}{#2}}}
\def\dfft#1#2{{\displaystyle\fft{#1}{#2}}}

\def\0{{\sst{(0)}}}
\def\1{{\sst{(1)}}}
\def\2{{\sst{(2)}}}
\def\3{{\sst{(3)}}}
\def\4{{\sst{(4)}}}
\def\5{{\sst{(5)}}}
\def\6{{\sst{(6)}}}
\def\7{{\sst{(7)}}}
\def\8{{\sst{(8)}}}
\def\sst#1{{\scriptscriptstyle #1}}

\def\ep{{\epsilon}}
\def\del{{\partial}}

\def\crampest{\medmuskip = 1mu plus 1mu minus 1mu}
\def\uncramp{\medmuskip = 4mu plus 2mu minus 4mu}

\def\u#1{{{\underline #1\,}}}

\def\cG{{{\cal G}}}
\def\cB{{{\cal B}}}

\def\im{{{\rm i\,}}}
\def\R{{\mathbb R}}

\def\bp{{{\bf\rm p}}}
\def\bb{{{\bf b}}}
\def\cQ{{{\cal Q}}}
\def\bZ{{{\bf Z}}}

\def\bmu{{{\bar\mu}}}
\def\bnu{{{\bar\nu}}}
\def\bphi{{{\bar\phi}}}

\def\scri{{\mathscr{I}}}

\def\rms{{{r_{\!\!\sst -}}}}
\def\rps{{{r_{\!\!\sst +}}}}

\def\half{\frac{1}{2}}
\def\ben{\begin{equation}}
\def\bea{\begin{eqnarray}}
\def\een{\end{equation}}
\def\eea{\end{eqnarray}}
\def \bp {{\bf p}}

\def \cG {{\cal  G}}

\def\ft#1#2{{\textstyle{\frac{\scriptstyle #1}{\scriptstyle #2} } }}
\def\fft#1#2{{\frac{#1}{#2}}}

\begin{document}

\begin{flushright}
\hfill {MI-HET-864
}\\
\end{flushright}

\begin{center}
{\large {\bf Perturbations of Black Holes
in  Einstein-Maxwell-Dilaton-Axion (EMDA) Theories
 }}

\vspace{10pt}
{\large  C.N. Pope$^{1,2}$, D.O. Rohrer$^{1}$ and  B.F. Whiting$^{3}$}

\vspace{10pt}

\hoch{1}{\it George P. \& Cynthia Woods Mitchell  Institute
for Fundamental Physics and Astronomy,\\
Texas A\&M University, College Station, TX 77843, USA}

\hoch{2}{\it DAMTP, Centre for Mathematical Sciences,
 Cambridge University,\\  Wilberforce Road, Cambridge CB3 OWA, UK}

\hoch{3}{\it Department of Physics, P.O. Box 118440, University of Florida,\\
Gainesville, FL 32611-8440, USA}

\vspace{5pt}



\end{center}

\begin{abstract}

{\normalsize

We extend our earlier work on the linearised perturbations of 
static black holes in Einstein-Maxwell-Dilaton (EMD) theories to the case 
where the black holes are solutions in an enlarged theory including
also an axion.  We study the perturbations in a 3-parameter family of
such EMDA theories.  
The systems of equations
describing the linearised perturbations can always be separated, but they
can only be decoupled when the three parameters are restricted to a 1-parameter
family of EMDA theories, characterised by a parameter $b$ that
determines the coupling
of the axion to the 
$\epsilon^{\mu\nu\rho\sigma}\, F_{\mu\nu}\, F_{\rho\sigma}$ term.  
In the specific case when $b=1$, the theory is related to
an ${\cal N}=2$ supergravity.  In this one case we find that the perturbations
in the axial and the polar sectors are related by a remarkable transformation,
which generalises one found by Chandrasekhar for the perturbations of
Reissner-Nordstr\"om in Einstein-Maxwell theory. 
This transformation is of a form found in supersymmetric
quantum mechanical models.  The existence of such 
mappings between the axial and polar perturbations appears to
correlate with those cases where there is an 
underlying supergravity supporting the solution,
even though the black hole backgrounds are non-extremal 
and therefore not supersymmetric.
We prove the mode stability of the static
black hole solutions in the supersymmetric EMDA theory.  For other
values of the parameter $b$ in the EMDA theories that allow decoupling 
of the modes, we find that one of 
the radial potentials can be negative outside the horizon if $b$ is 
sufficiently large,
raising the possibility of there being perturbative mode instabilities in
such a case.
}

\end{abstract}

{\small 
pope@physics.tamu.edu, rohrer@physics.tamu.edu, bernard@phys.ufl.edu }
\pagebreak

\tableofcontents
\addtocontents{toc}{\protect\setcounter{tocdepth}{2}}

\section{Introduction}

   The study of linearised perturbations of black hole solutions in
Einstein and Einstein-Maxwell theories has a long history, dating back
to the investigations of the perturbations of the Schwarzschild solution
of Einstein gravity or Einstein-Maxwell theory 
by Regge and Wheeler in the 1950s \cite{regwhe,wheeler}.  Subsequent work in
the 1970s and 1980s extended these results by considering black holes
carrying electric charge and/or angular momentum.  
An important application of these early perturbative analyses was that
they allowed one to initiate an investigation of stability.  Stability
analysis itself requires a multi-level approach, the first step of which
is always the addressing of the question of mode stability.  This is a
question that can sometimes (but only sometimes) be easily settled by an
energy integral analysis, such as that provided in \cite{wald2} for
Schwarzschild, in \cite{whiting} for Kerr, and by us in \cite{porowh} for
static EMD black holes.  We will find that an energy
integral analysis suffices to establish stability for some, but not all,
of the perturbations we discuss. 

  Our own work closely parallels that of Chandrasekhar and Xanthopoulos,
in their study of the linearised perturbations of Reissner-Nordstr\"om 
black holes in Einstein-Maxwell theory 
\cite{chandra,chanxant,xanth1,xanth2,chandra3}.  
This involved making 
gauge choices so as to isolate the perturbative degrees of freedom 
of the metric and Maxwell fields into quantities obeying 
a tractable system of coupled differential equations that could eventually
be separated and diagonalised.  One of the striking features that was
found in this analysis of Reissner-Nordstr\"om 
was that the axial modes and the polar modes,
which decouple from each other at the outset of the analysis, obey
ostensibly very different systems of second-order differential equations
which, nonetheless, turn out to be closely related to one another.  In
particular, the spectra of modes in the axial and polar sectors
are related by a one-to-one mapping, of the kind encountered in
supersymmetric quantum mechanical models.

  A rather different approach to the study of linearised black-hole
perturbations was developed by Teukolsky, who worked in the 
Newman-Penrose tetrad formulation of the Einstein or Einstein-Maxwell
equations, with the perturbative degrees of freedom being described
by gauge-invariant quantities involving the Weyl curvature and Maxwell
field strength.  In the case of the perturbations of the
Reissner-Norstr\"om black hole, Teukolsky's approach 
provides a unified description of
the axial and polar perturbations of the Chandrasekhar and Xanthopoulos
analysis in terms of complex fields obeying a single master equation. 

  The Teukolsky approach can also be applied to the case of rotating 
black holes, proving again a complete separation and decoupling of
the second-order perturbation equations.  However, no analogous 
complete diagonalisation of the equations has been achieved in the case
that both charge and rotation are turned on.

   In our previous work on the perturbations of static black holes
in the Einstein-Maxwell-Dilaton (EMD) theories, we were able to carry out
a complete separation and diagonalisation of the system of equations,
both in the axial and polar sectors \cite{porowh}.  It was evident, however,
that the close relation between the modes of the axial and polar
sectors that was previously seen in the perturbations of the
Reissner-Nordstr\"om solution in Einstein-Maxwell theory could no longer
occur in the EMD case.  Principally, in the EMD case the axial
perturbations are governed by two decoupled second-order equations,
while the polar perturbations are governed by three decoupled
second-order equations,
and there can no longer exist a one-to-one mapping between the modes in the
two sectors.  A natural expectation might therefore be that a Teukolsky
type of analysis of the perturbations could be problematical.  

   In the present paper we extend beyond the EMD system, by including now
in addition an axionic scalar field in the theory.  Specifically, the
class of EMDA theories that we shall consider is described by the
Lagrangian
\bea
{\cal L}= \sqrt{-g}\, \Big[ R -\ft12 (\del\phi)^2 - e^{a\phi}\, F^2
-\ft12 e^{c\phi}\,(\del\chi)^2 + 
   b\,\chi\,\widetilde{F}^{\mu\nu}F_{\mu\nu}\Big]\,,
\label{emdalag}
\eea
where $\widetilde{F}^{\mu\nu}\equiv\ft12 \,
         \ep^{\mu\nu\rho\sigma}F_{\rho\sigma}$
 is the Hodge dual of the Maxwell tensor $F_{\mu\nu}$.  The constants
$a$, $b$ and $c$ can {\it a priori} be taken to be arbitrary.  

   The background static solutions that we shall consider in these
EMDA theories will be exactly the same charged static 
Gibbons-Maeda black holes 
\cite{gibmae}
that served as the background solutions in the EMD theories 
\cite{porowh}.  In particular, the axion field vanishes in the
background solution.  However, the axion does enter at the level of the 
linearised perturbations
around the background solution, and in fact it gives rise to a third
degree of freedom obeying a second-order differential equation in
the axial sector.  This means that there will now be three perturbative
modes in each of the axial and polar sectors, raising the possibility 
that a matching of the axial and the polar modes might now occur.
Indeed, as we shall show, for appropriate choices  of the coupling
constants $a$, $b$ and $c$, we do find that the axial and polar modes
are related in an extension of the relations seen for the perturbations
of Reissner-Nordstr\"om in Einstein-Maxwell theory.

    In the EMD truncation of the EMDA system 
(in which the axion field $\chi$ is set to zero), the
Lagrangian (\ref{emdalag}) is invariant under a global symmetry
\bea
\phi\longrightarrow \phi + \lambda\,,\qquad
  A_\mu\longrightarrow e^{-\ft12 a\, \lambda}\, A_\mu\,.\label{Rsym1}
\eea
The kinetic term for the axion in the EMDA theory 
will share this symmetry too if the
axion is scaled also, according to
\bea
\chi\longrightarrow e^{-\ft12 c\lambda}\, \chi\,.\label{Rsym2}
\eea
However, the Chern-Simons term $b\,\chi\, \widetilde{F}^{\mu\nu}F_{\mu\nu}$
will break the global symmetry unless the constants $a$ and $c$ are chosen
such that
\bea
c=-2a\,.\label{carel}
\eea
For any choice of the parameters there is, of course, another
global symmetry under which the axion is shifted by an additive 
constant.  

In what follows we shall not, for now, impose the relation (\ref{carel}) but,
as will emerge later, the diagonalisation of the axial perturbation equations
seems only to be possible if eqn (\ref{carel}) holds, together with the
additional restriction that $a=1$.  

    In section 2 we set up the perturbative analysis, giving the background
solution and then obtaining the equations describing linearised perturbations
around the background.  We focus exclusively on the axial sector in this
section, because in the polar sector the linearised analysis is identical
to that in the EMD theories that we studied extensively in \cite{porowh}. 
As in our previous paper, we consider the full system of scalar, 
electromagnetic and gravitational perturbations.  All the decoupled equations
we eventually obtain contain contributions from the metric perturbations,
and hence are only applicable when the angular mode number $\ell$, which
arises in the separation of variables, is greater than or equal to 2.   

In section 3 we separate variables in the axial sector, obtaining
three coupled second-order radial equations.  For general values of the
parameters $a$, $b$ and $c$ in the EMDA Lagrangian it does not appear
to be possible to diagonalise the three second-order radial equations.  In
section 4 we follow a slightly different approach in which the linearised
axial perturbations are described in terms of a system of six coupled 
first-order radial equations, and then in section 5 we show how the
system can be diagonalised in the case where the parameters 
$a$ and $c$ in the EMDA Lagrangian are specialised to the values $a=1$, 
$c=-2$.  The axial perturbations can then be described in terms of
three decoupled second-order radial equations of the form
$(\del_{r_*}^2 +\omega^2 -V^-_{(p)} )\, Z^-_{(p)}=0$, where $p$ is
a constant that can be any of the three roots of a certain cubic
equation.

   In section 6 we show that if the remaining parameter $b$ in the 
specialised EMDA Lagrangian is taken to be $b=\pm1$, then the three potentials
$V^-_{(p)}(r)$ in the axial sector can be related, one-to-one, with 
the three potentials $V^+_{(p)}(r)$ that we found previously in
the polar sector in \cite{porowh}.  (In particular, in the $b^2=1$ 
case the constant $p$ in the axial sector then satisfies exactly the same 
cubic equation as the one arising in the polar sector.)  The relation
that we find between $V^+_{(p)}(r)$ and $V^-_{(p)}(r)$ for each of the
three $p$ values is of the same kind that was previously found
by Chandrasekhar in the simpler cases of Schwarzschild or Reissner-Nordstr\"om
black holes in Einstein or Einstein-Maxwell gravity, in which a pair of 
axial and
polar potentials $V^\pm$ can be written in terms of a superpotential 
$W=f+\beta$,
with $V^\pm = \pm \del_{r_*} f + f^2 + 2\beta\, f= 
\pm \del_{r_*} W + W^2-\beta^2$, where $\beta$ is 
a constant.  The solutions for the radial modes in the axial and polar 
sectors can then also be mapped into one another.  
Such systems are known in the literature as ``supersymmetric
quantum mechanical'' models (see, for example, 
\cite{witten1981,hayrau,valmorber,tong}).  

     We also show explicitly in section 6
that the axial potentials $V^-_{(p)}(r)$ are all non-negative outside
the outer horizon in the $b^2=1$ case.  An analogous result in the
polar sector was obtained previously in \cite{porowh}.  These results
are sufficient to establish mode stability for the charged static Gibbons-Maeda
black holes, viewed now as solutions in the $a=1$, $c=-2$, $b^2=1$
EMDA theory.

   A striking feature of supersymmetric quantum mechanical relations between
the axial and polar potentials in the Schwarzschild and Reissner-Nordstr\"om
cases studied by Chandrasekhar is that the theories in which the
black hole backgrounds are embedded can themselves be extended to
supergravities.  That is to say, the theory of Einstein gravity in
which the Schwarzschild perturbations were studied can be supersymmetrised
by adding a fermionic sector, giving ${\cal N}=1$ supergravity.  Likewise,
the Einstein-Maxwell theory in which the Reissner-Nordstr\"om perturbations
were analysed can be supersymmetrised to ${\cal N}=2$ supergravity.  It
is tempting to think that there may be a deeper underlying relation between
the ``supersymmetry'' of the supersymmetric quantum mechanical models
that arise in these perturbation analyses and the fact that the underlying 
theories in which the black holes reside are supergravities.

   Our new results in this paper may provide further insight into this 
question.  The supersymmetric quantum mechanical relation between the
axial and polar modes that we find here works {\it only} when the
parameter $b$ in the $a=1$, $c=-2$ family of EMDA Lagrangians takes the
value $b=1$ or $b=-1$.  Remarkably, it is precisely in these cases that
the EMDA theory is related to a supergravity theory.  In this case
the $a=1$, $c=-2$, $b=\pm1$ Lagrangian is not itself the bosonic sector
of a supergravity theory, but we can easily augment the Lagrangian by
adding an additional Maxwell field $H_{\mu\nu}$, to give the Lagrangian
(\ref{truncSTUlag}), and this theory is then the bosonic sector of
a supergravity (in fact, a consistent supersymmetric truncation of 
${\cal N}=2$ STU supergravity). If we wish to extend the notion of 
linking the supersymmetric quantum mechanical relation between
axial and polar modes to a possible underlying explanation rooted in
a supergravity, then in our current case we should also include the
perturbations of the $H_{\mu\nu}$ field in the analysis.  As can be
seen from the Lagrangian (\ref{truncSTUlag}), the charged static Gibbons-Maeda
black holes will still be solutions in this extended theory (with $H_{\mu\nu}=0$
in the background), and the analysis of the perturbations of $H_{\mu\nu}$
will be rather simple, because they will be totally decoupled from all
the other perturbations.  We carry out this analysis also in section 6, 
and show that indeed the axial and polar perturbations in the 
$H_{\mu\nu}$ sector have potentials that are again related via a 
supersymmetric 
quantum mechanical model; in this case, a very simple one.  Thus the
notion of a linkage between the occurrence of the supersymmetric
quantum mechanical relations on the one hand, and the fact that the
underlying theories where this arises are supergravities, is somewhat 
strengthened by our results.

   In section 7 we consider the perturbations in the more general
one-parameter family of EMDA theories where $a=1$ and $c=-2$ but
with $b$ now arbitrary.  The polar sector was already fully analysed
in \cite{porowh} (now specialised to $a=1$), and in particular the 
polar potentials were seen to be always non-negative outside the horizon.
The situation with the axial potentials obtained in this paper turns out
to be more interesting.  We find that of the three axial potentials
$V^-_{(p)}(r)$, where $p$ is any of the three roots of the cubic
equation (\ref{pcubic}), only the case when $p$ is positive (it is always
the case that just one root is positive) has the possibility of
having the potential be negative outside the outer horizon.  We find the
conditions under which this can in fact occur, and establish lower 
bounds on the value of $b^2$, beyond which there can be a region outside the
horizon for which the potential is negative.  The lowest positive 
value for which this can occur is for the minimum angular mode number
$\ell=2$, and the bound is $b>\sqrt{\dfft{365}{112}} = 1.80525\ldots$.
Thus if the coupling of the axion to the topological term
$\epsilon^{\mu\nu\rho\sigma}\, F_{\mu\nu}\, F_{\rho\sigma}$ is too large,
then the possibility of mode instabilities can arise.

  The paper concludes with a summary and further discussion in section 8. 
Included in this discussion is an elaboration of the relations between
the axial and polar sectors in the $a=1$, $c=-2$, $b^2=1$ theories and the
general notation of supersymmetric quantum mechanical models.  There are also
two appendices.  In appendix A, we give a more detailed discussion of the
system of three coupled second-order radial equations in the $a=1$, $c=-2$
theories, and their relation to the diagonalised equations we obtained 
in section 5.  In appendix B we exhibit some plots of the various
axial and polar potentials for different values of $b$, showing in particular
how one of the axial potentials can become negative outside the outer horizon
when $b$ is sufficiently large.

\section{The Linearised Perturbations}

\subsection{The background solutions}

 The equations of motion following from the Lagrangian (\ref{emdalag}) 
are
\bea
R_{\mu\nu} &=& \ft12 \del_\mu\phi\, \del_\nu\phi + \ft12 e^{c\phi}\del_\mu\chi\,
 \del_\nu\chi + 2 e^{a\phi}\,
   (F_{\mu\rho}\, F_\nu{}^\rho - \ft14 F^{\rho\sigma}\, F_{\rho\sigma}\,
g_{\mu\nu})\,,\nn\\
\nabla_\nu (e^{a\phi}\, F^{\mu\nu} - b\, \chi\, \widetilde{F}^{\mu\nu}) &=&0\,,
\nn\\
\square\phi &=& a\, e^{a\phi}\, F^2 + \ft12 c\,e^{c\phi}\,(\del\chi)^2\,,\nn\\
\nabla_\mu(e^{c\phi}\,\nabla^\mu\,\chi) + b\,\widetilde{F}^{\mu\nu}F_{\mu\nu} &=
& 0\,.
\label{emdaeom}
\eea

 The static electrically-charged black hole solutions of the EMD theory
were constructed by Gibbons and Maeda \cite{gibmae}.  
In the notation we shall be using here, which is the same as in \cite{porowh},
as solutions of the EMDA theory (\ref{emdalag}) they are given by
\bea
ds^2 &=& -\Delta\, dt^2 + \fft{dr^2}{\Delta} + R^2\, (d\theta^2 +
\sin^2\theta\, d\varphi^2)\,,\nn\\
e^{-a\phi} &=& f_-^{\ft{2a^2}{1+a^2}}\,,\qquad \chi=0\,,\qquad
\qquad A= \fft{Q}{r}\, dt\,,\nn\\
\Delta &=& f_+\, f_-^{\ft{1-a^2}{1+a^2}}\,,\qquad 
R = r\, f_-^{\ft{a^2}{1+a^2}}\,,\qquad f_\pm = 1-\fft{r_\pm}{r}\,.
\label{bhsol}
\eea
(Note that we are now including the axion $\chi$, which vanishes in these
background solutions.) 
The electric charge $Q$ is related to the radii $r_\pm$ of the outer
and inner horizons by
\bea
Q^2= \fft{\rps\, \rms}{1+a^2}\,.
\eea
Note that in the limit when the dilaton coupling $a$ goes to zero, the
metric and Maxwell field reduce to those of the Reissner-Nordstr\"om
solution, and the dilaton field vanishes.  

\subsection{Parameterisation of the perturbations}

  Since the background is 
static and spherically symmetric, the perturbations of the metric can, 
without loss of generality, be
taken to have the form
\bea
ds^2 = -e^{2\nu}\, dt^2 + e^{2\mu_1}\, dr^2 + e^{2\mu_2}\, d\theta^2 +
   e^{2\psi}\, (d\varphi - \cB_\alpha \, dx^\alpha)^2\,,\label{genmet}
\eea
where we index the coordinates $(t,r,\theta,\varphi)$ as $(0,1,2,3)$, the
index $\alpha$ ranges over
$\alpha=(0,1,2)$, and all quantities are taken to depend on
$t$, $r$ and $\theta$, but to be independent of
the azimuthal angle $\varphi$.  Note that the metric has the form of
a Kaluza-Klein reduction from 4 to 3 dimensions, where the 
reduction is performed on the 
azimuthal coordinate $\varphi$.  

    We shall choose the natural tetrad frame $e^{\u a}
=e^{\u a}{}_\mu\, dx^\mu$ with
\bea
e^\u0= e^\nu\, dt\,,\qquad e^\u1= e^{\mu_1}\, dr \,,
  \qquad e^\u2= e^{\mu_2}\, d\theta\,,\qquad 
e^\u3= e^{\psi}\, (d\varphi -\cB_\alpha \, dx^\alpha)
\eea
in order to describe the metric (\ref{genmet}).
Note that we are using the index $\u a$ 
to run over the tetrad-frame values
$(\u0,\u1,\u2,\u3)$ and the index $\mu$ to run over the coordinate-frame
values $(0,1,2,3)$.  
The inverse tetrad $E_{\u a}{}^\mu$ is
given by
\bea
E_\u0{}^\mu\del_\mu &=& e^{-\nu}\,\Big(\fft{\del}{\del t} + \cB_0\,
\fft{\del}{\del\varphi}\Big)\,,\qquad
E_\u1{}^\mu\del_\mu= e^{-\mu_1}\, \Big(\fft{\del}{\del r} + \cB_1\,
\fft{\del}{\del\varphi}\Big)\,,\nn\\
E_\u2{}^\mu\del_\mu &=& e^{-\mu_2}\, \Big(\fft{\del}{\del \theta} + \cB_2\,
\fft{\del}{\del\varphi}\Big)\,,\qquad
E_\u3{}^\mu\, \del_\mu = e^{-\psi}\, \fft{\del}{\del\varphi}\,.
\eea
 
The complete set of fields parameterising configurations of the
perturbed system comprise 
the metric fluctuations $(\delta\nu,\delta\mu_1,\delta\mu_2, \delta\psi,
\delta\cB_\alpha)$, the Maxwell field fluctuations 
$\delta F_{\u a \u b}$,
and the dilaton and axion fluctuations $\delta\phi$ and $\delta\chi$.  
As has been mentioned, these can be divided into the two disjoint sets
\bea
\hbox{Axial}:&& (\delta\cB_0\,,\delta\cB_1\,,\delta \cB_2\,, \delta
 F_{\u0\u3}\,,
\delta F_{\u1\u3}\,,\delta F_{\u2\u3}\,,\delta\chi)\,,\label{axiallist}\\
\hbox{Polar}:&& (\delta\nu\,,\delta\mu_1\,,\delta\mu_2\,,\delta\psi\,,
\delta F_{\u0\u1}\,,\delta F_{\u0\u2}\,,\delta F_{\u1\u2}\,,\delta\phi)\,,
\label{polarlist}
\eea
with the axial fluctuations changing sign under a reversal of the azimuthal
angle $\varphi$, while the polar fluctuations remain invariant.  Because the
background solution is azimuthally invariant, the equations governing the
linearised axial fluctuations will then necessarily be decoupled 
from those governing the polar fluctuations.

   Note that since the axion does not enter into the linearised polar equations,
and because the axion vanishes in the background solutions, the analysis
of the polar sector will be completely unchanged from that given in
our previous paper \cite{porowh}, and so we can just refer to that paper for
all the polar results that we shall need.

   We now expand all the fields around the values that they
take in the Gibbons-Maeda black hole background, writing
\bea
\nu&=&\bar\nu + \ep\, \delta\nu\,,\qquad \mu_1=\bar\mu_1 + \ep\, \delta\mu_1
\,,\qquad \mu_2=\bar\mu_2 + \ep\, \delta \mu_2\,,\qquad
\psi=\bar\psi +\ep\, \delta\psi\,,\nn\\
\cB_\alpha &=& \ep\, \delta\cB_\alpha\,,\qquad
\phi =\bar\phi + \ep\, \delta\phi\,,\qquad \chi=\ep\, \delta\chi \,,
\label{metricpert}
\eea
together with the field strength
\bea
F_{\u a\u b}= \bar F_{\u a \u b} + \ep\, \delta F_{\u a \u b}\,.\label{maxpert}
\eea
By comparing  these expressions, together with eqn (\ref{genmet}) 
with the black hole solution in eqns (\ref{bhsol}), it follows that the
expressions for the background fields (denoted by the overbars) are
\bea
\bar\nu &=&\ft12\log\Delta\,,\qquad \bar\mu_1=-\bar\nu\,,\qquad
\bar\mu_2=\log R\,,\qquad \bar\psi=\log R+\log\sin\theta\,,\nn\\
\bar\cB_\alpha &=&0\,,\qquad
 a \bar\phi =2\log r-2\log R\,,\qquad \bar F_{\u0\u1} =\fft{Q}{r^2} 
\label{backgroundrels}
\eea
where $\Delta$ and $R$ are given in eqns (\ref{bhsol}), and all
components of the background field strength $\bar F_{\u a\u b}$ other than 
$ \bar F_{\u0\u1} =- \bar F_{\u1\u0}$ vanish. 

\subsection{The axial perturbation equations}

  In the tetrad basis that we are employing, the equations of motion
(\ref{emdaeom}) take the form
\bea
(\hbox{Ein})_{\u a\u b} &\equiv& R_{\u a\u b} 
-\ft12 R\, \eta_{\u a\u b}  -\ft12 \phi_\u a\, \phi_\u b 
+\ft14 \phi^{\u c}\,\phi_{\u c} \,\eta_{\u a\u b}
-\ft12 e^{c\phi} \chi_{\u a}\, \chi_{\u b} 
+\ft14 e^{c\phi}\,\chi^{\u c}\,\chi_{\u c}\, \eta_{\u a\u b}
\nn\\
&& - 2e^{a\phi}\, (F_{\u a\u c}\, F_{\u b}{}^{\u c} 
- \ft14 \eta_{\u a\u b} \, F^{\u c\u d} F_{\u c\u d})=0\,,
\label{tetradein}\\
(\hbox{Max})^{\u a} &\equiv& e^{\u a}{}_\mu\, 
\del_\nu\big(\sqrt{g}\, E^{\u b\mu}\, 
E^{\u c\nu}\,( \, e^{a\phi}\,F_{\u b\u c}- b\, \chi\,
\widetilde{F}_{\u b\u c})\big)=0\,,
\label{tetradmax}\\
E_\phi&\equiv & \square\phi-a\, e^{a\phi}\, F^{\u a\u b}\, F_{\u a\u b} 
-\ft12 c\, 
e^{c\phi}\,\chi^{\u a}\,\chi_{\u a}=0\,,\label{dileqn}\\
E_\chi &\equiv& \del_{\mu}(\sqrt{g}\,e^{c\phi}\,E^{\u a\mu}\chi_{\u a}) + 
             b\sqrt{g}\,\widetilde{F}^{\u a\u b}F_{\u a\u b}=0\,,\label{axeqn}
\eea
where $\phi_{\u a}=E^\mu_{\u a}\, \del_\mu\phi$, etc., and 
$\square\phi=\dfrac{1}{\sqrt{g}}\del_\mu(\sqrt{g}\, E^{\u a\mu}\,\phi_{\u a})$ 
in terms of tetrad quantities.  The Bianchi identities for the Maxwell
field are given by
\bea
(\hbox{Bianchi})^{\u a} \equiv e^{\u a}{}_\mu\, 
 \varepsilon^{\mu\nu\rho\sigma}\,
\del_\nu(e^{\u b}{}_\rho\, e^{\u c}{}_\sigma\, F_{\u b\u c})=0\,.
\label{tetradbianchi}
\eea

   The linearised axial equations of motion, involving the fields
listed in eqn (\ref{axiallist}), come from the components 
$(\hbox{Ein})_{\u1\u3}$ and $(\hbox{Ein})_{\u2\u3}$ of the Einstein equations
(\ref{tetradein}); the component $(\hbox{Max})^\u3$ of the Maxwell
equations (\ref{tetradmax}); the axion equation $E_\chi$ (\ref{axeqn});
and the components $(\hbox{Bianchi})^\u1$ and $(\hbox{Bianchi})^\u2$ of
the Bianchi identities (\ref{tetradbianchi}).  Since the only
non-vanishing component of the background Maxwell field $\bar F_{\u a\u b}$
is given by
$\bar F_{\u0\u1}= \dfft{Q}{r^2}$, it follows that the equations for the
linearised axial perturbations are as follows:  

From the Einstein equations 
(\ref{tetradein}) we have
\bea
(\hbox{Ein})_{\u1\u3}:&& e^{\bar\nu-\bar\mu_2}\, \fft1{\sin^2\theta}\,
\del_\theta(\sin^3\theta\, \delta \cG_{12}) + e^{-\bar\nu+\bar\mu_2}\,
\sin\theta\, \del_t\,\delta \cG_{01} - \fft{4Q}{r^2}\, 
  e^{a\bar\phi}\,
\delta F_{\u0\u3}=0\,,\nn\\
(\hbox{Ein})_{\u2\u3}:&& e^{-2\bar\mu_2}\,
\del_r(e^{2\bar\nu+2\bar\mu_2}\, \delta\cG_{12}) -e^{-2\bar\nu}\, 
 \del_t\,\delta \cG_{02}=0\,.\label{Ein1323}
\eea
(Note that $\cG=d\cB$ is the field strength of the Kaluza-Klein 1-form
$\cB$ appearing in the metric (\ref{genmet}), with coordinate components
$\cG_{\alpha\beta}$, with $\alpha$ and $\beta$ ranging over 0, 1 and 2. 
Following our
conventions the non-underlined numerical 
indices on $\cG$ in eqns (\ref{Ein1323}) indicate coordinate index values.)

The Maxwell equation (\ref{tetradmax}) gives
\bea
(\hbox{Max})^\u3:&&  e^{-\bar\nu-\bar\mu_2}\, \del_t\, \delta F_{\u0\u3} -
e^{-2\bar\mu_2-a\bar\phi}\, 
 \del_r(e^{\bar\nu+\bar\mu_2 +a\bar\phi}\, \delta F_{\u1\u3}) -\nn\\
&&e^{-2\bar\mu_2}\, \del_\theta 
\Big(\delta F_{\u2\u3}+ e^{-a\,\bar\phi}\, \fft{b\,Q}{r^2}\,\delta\chi\Big)
+ \fft{Q\,\sin\theta}{r^2}\, \delta\cG_{01}=0\,.\label{Max3}
\eea

The axion equation (\ref{axeqn}) gives
\bea
E_\chi:&& 
 -e^{-2\bnu+c\bphi}\,\del_t^2\,\delta\chi + e^{-2\bmu_2}\,\del_r(e^{2\bnu+2\bmu_2+c\bphi}\,\del_r\,\delta\chi) \nn\\ &&+e^{-2\bmu_2+c\bphi}\,\fft1{\sin\theta}\del_\theta(\sin\theta\,\del_\theta\,\delta\chi)-\fft{4b\,Q}{r^2}\delta F_{\u2\u3}=0\,.\nn\\\label{Ax}
\eea

Finally, the Bianchi identities (\ref{tetradbianchi}) give
\bea
(\hbox{Bianchi})^\u1:&&  \del_t\,\delta F_{\u2\u3}= 
  e^{\bar\nu -\bar\mu_2}\, \fft1{\sin\theta}\, 
  \del_\theta(\sin\theta\, \delta F_{\u0\u3})\,,\nn\\
(\hbox{Bianchi})^\u2:&&  \del_t\,\delta F_{\u1\u3}=
  e^{\bar\nu-\bar\mu_2}\, \del_r(e^{\bar\nu+\bar\mu_2}\, 
\delta F_{\u0\u3})\,,\label{B1B2M3}
\eea
Note that aside from the axion equation (\ref{Ax}), all the other 
equations (absent their axion contributions) are exactly the same as those
obtained in ref.~\cite{porowh}, where some more details of their derivation
can be found.

\section{\label{secondordersec} Second-Order Equations For Axial Perturbations}
 
  One approach to extracting the complete system of equations for the
axial perturbations is to construct directly a set of second-order
differential equations.  This follows essentially the same procedure
that was employed in \cite{chanxant} for Reissner-Nordstr\"om, and that
we employed in \cite{porowh} for the static charged black holes in
the EMD theories; now, of course, with the added complication of the
axion field.   

   An essential step, when analysing the system of axial equations, is
that we decompose all fluctuations into modes with time dependence 
$e^{-\im\omega t}$, with $\omega$ assumed to be non-zero.  Thus 
time derivatives in the equations of motion become
multiplication by $-\im\omega$, allowing certain fields to be 
solved for algebraically.  The general strategy is then, as set out in
\cite{porowh}, as follows:

  Taking the time derivative of $(\hbox{Max})^\u3$ and then using 
$(\hbox{Bianchi})^\u1$ and $(\hbox{Bianchi})^\u2$ allows $\delta F_{\u1\u3}$
and $\delta F_{\u2\u3}$ to be eliminated. $\delta \cG_{01}$ can then
be eliminated using $(\hbox{Ein})_{\u1\u3}$.   Another disentanglement of
equations can be achieved by taking the $\del_\theta$ derivative of
$(\hbox{Ein})_{\u1\u3}$, and the $\del_r$ derivative of $(\hbox{Ein})_{\u2\u3}$,
and using the Bianchi identity $\del_r\,\delta \cG_{02} -
\del_\theta\, \delta \cG_{01}= \delta_t\, \delta\cG_{12}$.  The upshot is 
that the equations can now be separated, yielding ODEs in $r$,
by writing
\bea
\delta F_{\u0\u3}(t,r,\theta) &=& \frac{\im\omega\, \ell(\ell+1)\, 
e^{-\bar\nu}}{r \sin\theta}\, e^{-\im\omega t}\,
B_{\u0\u3}(r)\,C_{\ell+1}^{-\frac12}(\cos\theta)\\ 
&=& \frac{-\im\omega \, e^{-\bar\nu}}{r}\,
e^{-\im\omega t}\,B_{\u0\u3}(r)\,\del_{\theta}P_\ell(\cos\theta) \nn\\\nn\\
\delta\cG_{12}(t,r,\theta) &=& \frac{2\im\omega\,\mu\,\ell(\ell+1)\, 
e^{-2\bar\nu-\bar\mu_2}}{3\sin^3\theta}\,
e^{-\im\omega t}\,\cG_{12}(r)\,C_{\ell+2}^{-\frac32}(\cos\theta)\nn\\
&=& \frac{-2\im\omega\, e^{-2\bar\nu-\bar\mu_2}}{\mu}\, e^{-\im\omega t}
\,\cG_{12}(r)\,\del_\theta\big(\csc\theta\,\del_\theta 
      P_\ell(\cos\theta)\big)\,,\label{B03G12sep}\\
\delta\chi(t,r,\theta) &=& -2\sqrt{\ell(\ell+1)}\, 
e^{-\bar\mu_2-\ft12 c \bar\phi}
\,e^{-\im\omega t}\,\chi(r)\,P_\ell(\cos\theta)
\,,\label{axionsep}
\eea
where we have defined  
\bea
\mu=\sqrt{\ell(\ell+1)-2}\label{mudef}
\eea
(recall that we are restricting the mode number $\ell$ to $\ell\ge 2$,
since we are considering the coupled perturbations that include the
gravitational modes). 
The quantities $C_n^{(\lambda)}(x)$ are the Gegenbauer 
polynomials.  (See eqn (139) of chapter 5 of \cite{chandra} for 
some useful recurrence relations for the Gegenbauer polynomials.  Note
that $C_\ell^{\frac12}(x)=P_\ell(x)$ where $P_\ell(x)$ are the Legendre
polynomials.)
Note that the Bianchi identity 
$(\hbox{Bianchi})^\u1$ can be used to obtain the algebraic relation
\bea
B_{\u2\u3}=-\ell(\ell+1)B_{\u0\u3}\,.
\eea

   Implementing the above procedures, we arrive at the following 
three coupled ordinary differential equations for the three radial functions
$B_{\u0\u3}(r)$, $\cG_{12}(r)$ and $\chi(r)$:
\bea
(\del_{r_*}^2 +\omega^2)\,  B_{\u0\u3} +T_1\,B_{\u0\u3} + S_1\,\cG_{12} +P_1\,\chi &=& 0\,,\nn\\
(\del_{r_*}^2+\omega^2)\, \cG_{12} + T_2\,\cG_{12} +
   S_1\, B_{\u0\u3} &=& 0\,,\nn\\
(\del_{r_*}^2+\omega^2)\, \chi+ T_3\,\chi + P_1\,B_{\u0\u3} &=& 0
\,,\label{secondorderAxialeqs}
\eea
where
\bea
T_1 &=& -e^{4\bnu}\, \Big(a\,\bnu'\,\bphi{'} + \ft14 a^2\, {\bphi{'}}^2 + \ft12 a\,\bphi''\Big) -\fft1{r^2}\, e^{2\bnu+a\bphi}\,\Big(\fft{4Q^2}{r^2} +\ell(\ell+1)\Big)\,,\nn\\
T_2 &=& - e^{4\bnu} \,\Big(a \,\bnu'\,\bphi' + \ft14 a^2\, {\bphi{'}}^2 + \ft12 a\,\bphi'' + \fft{2}{r^2}-\fft{2}{r}\, \bnu' - \fft{a}{r}\, \bphi' \Big) - \fft1{r^2}\, e^{2\bnu+a\bphi}\,\Big(\ell(\ell+1) -2\Big)\,,\nn\\
T_3 &=& -e^{4\bnu}\, \Big(-(a-c)\,\bnu'\,\bphi' + \ft14 (a-c)^2\, {\bphi{'}}^2 - \ft12 (a-c)\,\bphi'' + \fft2r\bnu' -\fft{(a-c)}{r}\bphi'\Big) \nn\\
&&-\fft1{r^2}\,e^{2\bnu+a\bphi}\,\ell(\ell+1)\,,\nn\\
S_1 &=& -\fft{2\mu\,Q}{r^3} e^{2\bnu+a\bphi}\,,\nn\\
P_1 &=& -\fft{2b\,\sqrt{\ell(\ell+1)}\, Q}{r^3} \,
   e^{2\bnu -\ft{c}{2}\bphi}\,.\label{tspcoefficients}
\eea
Note that the derivatives $\del_{r_*}$ in eqns (\ref{secondorderAxialeqs})
are with respect to the tortoise coordinate $r_*$ defined by
\bea
\del_{r_*}= e^{2\bar\nu}\, \del_r\,.
\eea
We have also made use of the background relations given in eqns 
(\ref{backgroundrels}) to write $\bar\mu_2$ in terms of $r$ and $\bar\phi$
using
\bea
e^{\bar\mu_2}= r\, e^{-\ft12 a \bar\phi}\,.
\eea

  For general values of the coupling constants $a$, $b$ and $c$ in the EMDA 
theory (\ref{emdalag}) it does not appear to be possible to 
diagonalise the three second-order radial equations 
(\ref{secondorderAxialeqs}).  We have, however, succeeded in diagonalising the
equations in the special case when 
\bea
a=1\,,\qquad c=-2\,,\label{acspec}
\eea
Notice that this is
a particular instance of the $c=-2a$ class of EMDA theories that
we mentioned in the introduction, for which the Lagrangian has 
the global $\R$ symmetry given in eqns (\ref{Rsym1}) and (\ref{Rsym2}). 

\section{First-Order Equations For Axial Perturbations}

   It will be useful at this stage to adopt an approach that rather
closely parallels the discussion of the polar sector that we gave 
previously in \cite{porowh}, as this will be helpful later on when we find
circumstances under which the axial and polar sectors can be related. This
approach is based on directly dealing with a system of coupled
first-order radial equations.
We can obtain our system of separated first-order equations by 
making the following substitutions in the equations (\ref{Ein1323}),
(\ref{Max3}), (\ref{Ax}) and (\ref{B1B2M3}): 
\bea
\delta\cG_{01} &=& 2\,e^{-\bmu_2}\,e^{-\im\omega t}\,\cG_{01}(r)\,\csc\theta\,\del_\theta P_\ell\,,\nn\\
\delta\cG_{02} &=& \fft{2\,e^{-\bmu_2}}{\mu}\,e^{-\im\omega t}\,\cG_{02}(r)\,\del_\theta(\csc\theta\,\del_\theta P_\ell)\,,\nn\\
\delta\cG_{12} &=& -2\,\im\omega\, \frac{e^{-2\bar\nu-\bar\mu_2}}{\mu}\,e^{-\im\omega t}\,\cG_{12}(r)\,\del_\theta(\csc\theta\,\del_\theta P_\ell)\,,\nn\\
\delta F_{\u0\u3} &=& -\im\omega\,\frac{e^{-\bar\nu}}{r}\,e^{-\im\omega t}\,B_{\u0\u3}(r)\,\del_{\theta}P_\ell\,, \nn\\
\delta F_{\u1\u3} &=& \fft{e^{-\bnu}}r\,e^{-\im\omega t}\,B_{\u1\u3}(r)\,\del_{\theta}P_\ell\,, \nn\\
\delta F_{\u2\u3} &=& \frac{e^{-\bmu_2}}{r}\,e^{-\im\omega t}\,B_{\u2\u3}(r)\,P_{\ell}\,,\nn\\
\delta\chi &=& -2\sqrt{\ell(\ell+1)}\,e^{-\bar\mu_2-\ft12 c \bar\phi}\,e^{-\im\omega t}\,\chi(r)\,P_\ell\,.\label{axsepsV2}
\eea
Note that the fields $\cG_{12}$, $B_{\u0\u3}$ and $\chi$ are the ones that 
remained in
our previous construction of the second-order equations 
(\ref{secondorderAxialeqs}), after making use of the various Bianchi identities,
Maxwell and Einstein equations. Now, by no longer
using the Bianchi, Maxwell and Einstein equations to eliminate variables,
we can recast the system as an extended set of variables obeying
first-order equations, as detailed below:

Substituting eqns (\ref{axsepsV2}) into the Bianchi, Maxwell, Einstein 
and axion equations, the angular dependence can be seen to 
factor out upon using standard properties of the
Legendre polynomials, as does the time dependence, and we are left with the
purely radial equations
\bea
(\hbox{Ein})_{\u1\u3}:&& \fft{2Q}{r}\,B_{\u0\u3}-e^{2\bmu_2}\,\cG_{01}+\sqrt{(\ell+2)(\ell-1)}\,\cG_{12}=0\,,\label{rein13}\\
(\hbox{Ein})_{\u2\u3}:&& e^{-\bmu_2}\,\del_r\big(e^{\bmu_2}\,\cG_{12}\big)-e^{-2\bnu}\,\cG_{02}=0\,,\label{rein23}\\
(\hbox{Bianchi})^{\u1}:&&
\ell(\ell+1)\,B_{\u0\u3} + B_{\u2\u3}=0\,,\label{rbianchi1}\\
(\hbox{Bianchi})^{\u2}:&& 
r\,e^{-\bmu_2}\,\del_r\Big(\fft{1}{r}\,e^{\bmu_2}\,  B_{\u0\u3}\Big) - e^{-2\bnu}\,B_{\u1\u3}=0\,,\label{rbianchi2}\\
(\hbox{Max})^{\u3}:&&  \fft{1}{r}\,e^{\bmu_2}\,\del_r\big(r\,e^{-\bmu_2}\,B_{\u1\u3} \big)+\omega^2\,e^{-2\bnu}\,B_{\u0\u3}+e^{-2\bmu_2}\,B_{\u2\u3}\nn\\
&&-\fft{2Q}{r}\,\cG_{01}-\fft{2b\sqrt{\ell(\ell+1)}\,Q}{r^3}\,e^{-\ft12 c\bphi}\,\chi =0\,,\label{rmax3}\\
E_\chi:&& \chi'' + 2\bnu'\,\chi' + \big[\omega^2\,e^{-4\bnu}-\bmu_2'(\bmu_2'+2\bnu')-c\,\bphi'(\bmu_2'+\bnu'+\ft14c\bphi')-\bmu_2''\nn\\
&&-\ft12 c\,\bphi''-\ell(\ell+1)\,e^{-2\bmu_2-2\bnu}\big]\,\chi + \fft{2b\,Q}{\sqrt{\ell(\ell+1)}}\fft{e^{-2\bnu-\ft12 c\bphi}}{r^3}\,B_{\u2\u3}
 =0\,,\label{raxeom}\\
d\cG:&& e^{\bmu_2}\,\del_r\big(e^{-\bmu_2}\,\cG_{02}\big)+ 
\omega^2 \,e^{-2\bnu}\,\cG_{12}-\sqrt{(\ell+2)(\ell-1)}\,\cG_{01} =0
\,.\label{rdG}
\eea
(We have used the background relations in reducing the equations to this form.)

We now introduce the auxiliary variable $\xi=\chi'$, and we also make use 
of the purely algebraic equations 
(\ref{rein13}) and (\ref{rbianchi1}) to solve
for $\cG_{01}$ and 
$B_{\u2\u3}$, thus eliminating them from the remaining system of
differential equations. 
We are left with six independent variables:
\bea
\{\cG_{02},\cG_{12},B_{\u0\u3},B_{\u1\u3},\chi,\xi\}\,,\label{sixvar}
\eea
which will satisfy first-order equations.
As in the Reissner-Nordstr\"om  and EMD cases we define 
\bea
n=\ft12\ell(\ell+1)-1 =\ft12 \mu^2\,. \label{ndef}
\eea
 
    Thus we arrive at the system of six first-order equations
\bea
\cG_{02}' &=& \bmu_2'\,\cG_{02}+\big(2n\,e^{-2\bmu_2}-\omega^2\,e^{-2\bnu}\big)\,\cG_{12}+\fft{2\sqrt{2n}\,\,Q}{r}\,e^{-2\bmu_2}\,B_{\u0\u3} \,,\nn\\
\cG_{12}' &=& e^{-2\bnu}\,\cG_{02}-\bmu_2'\,\cG_{12} \,,\nn\\
B_{\u0\u3}' &=& \Big(\fft{1}r-\bmu_2'\Big)\,B_{\u0\u3}+e^{-2\bnu}\,B_{\u1\u3} \,,\nn\\
B_{\u1\u3}' &=& \Big(2(n+1)\,e^{-2\bmu_2}-\omega^2\,e^{-2\bnu}+\fft{4Q^2}{r^2}\,e^{-2\bmu_2}\Big)\,B_{\u0\u3}-\Big(\fft{1}{r}-\bmu_2'\Big)\,B_{\u1\u3}\nn\\
&&+ \fft{2\sqrt{2n}\,\,Q}{r}\,e^{-2\bmu_2}\,\cG_{12} + \fft{2b\sqrt{2(n+1)}\,Q}{r^3}\,e^{-\ft12 c\bphi}\,\chi \,,\nn\\
\chi' &=& \xi \,,\nn\\
\xi' &=& \big[2(n+1)\,e^{-2\bmu_2-2\bnu}+\bmu_2'(\bmu_2'+2\bnu')+c\,\bphi'(\bmu_2'+\bnu'+\ft14 c\bphi')+\bmu_2''+\ft12 c\,\bphi''-\omega^2\,e^{-4\bnu}\big]\,\chi \nn\\ 
&& +\fft{2b\sqrt{2(n+1)}\,Q}{r^3}\,e^{-2\bnu-\ft12 c\bphi}\,B_{\u0\u3}-2\bnu'\,\xi \,.\label{6firstorder}
\eea

\section{Diagonalisation Of The Axial Modes\label{diagsec}}

   Having obtained the system of six coupled first-order equations
(\ref{6firstorder}) describing the axial perturbations of the 
static black holes in the EMDA theories (\ref{emdalag}), 
we now proceed by using these to derive diagonalised second-order equations,
of the form
\bea
(\del_{r_*}^2+\omega^2)Z^- - V^-\,Z^- = 0\,,\label{zwaveeq}
\eea
where, for each $Z^-$, the function $V^-(r)$ will be the potential.
The fact that the perturbations in the axial sector can be completely 
described by three coupled second-order equations for 
$B_{\u0\u3}$, $\cG_{12}$ and $\chi$, as we saw in section 
\ref{secondordersec}, 
suggests that we should try an ansatz of the form
\bea
Z^- = f_1\,\cG_{12}+  f_2\,B_{\u0\u3}+ f_3\,\chi\,,\label{zminus} 
\eea
where the functions $f_i(r)$ are to be determined.

 Substituting (\ref{zminus}) into (\ref{zwaveeq}) and making 
repeated use of the six first-order equations (\ref{6firstorder}) yields
an equation of the form
\bea
e_1\,\cG_{02}+ e_2\,\cG_{12}+ e_3\,B_{\u0\u3}+ e_4\,B_{\u1\u3}+ 
e_5\,\chi+ e_6\,\xi=0\,.\label{e1to6equation}
\eea
Since the six fields $(\cG_{02},\cG_{12},B_{\u0\u3},B_{\u1\u3}, \chi, \xi)$
are treated as being independent here, this means that the coefficients
$e_j$ must all vanish,
\bea
e_j=0\,,\qquad 1 \leq j \leq 6\,.\label{eieqns}
\eea
Each of the coefficients $e_j$ will depend upon the functions $f_i$, the
background metric functions, and derivatives thereof. The coefficients
$e_2$, $e_3$ and $e_5$ will also depend on the unknown potential
function $V^-$.

  The coefficients $e_1$, $e_4$, and $e_6$ are independent of $V^-$, 
and each consists of a single term that can immediately be solved for 
$f_1$, $f_2$, and $f_3$. In fact, we simply have
\bea
e_1=2f_+f_-^{\fft{1-a^2}{1+a^2}}\,f_1'=0 &\implies& f_1=c_1\,,\nn\\
e_4=2f_+f_-^{\fft{1-a^2}{1+a^2}}\, f_2'=0 &\implies& f_2= c_2\,,\nn\\
e_6=2(f_+f_-)^2f_-^{\fft{-4a^2}{1+a^2}}\, f_3'=0 &\implies& f_3=c_3\,,
\label{tf123}
\eea
for constants $c_i$ to be determined.
Substituting these constant expressions for the $f_i$ 
back into eqns (\ref{e1to6equation}), the remaining equations for
$j=2, 3$ and 5 lead to the following conclusions:

    First we solve $e_2=0$ for the potential $V^-$, finding
\bea
V^- &=& \fft{1}{(1+a^2)^2\, c_1\,r^5}\,
 f_+ f_-^{\ft{-4a^2}{1+a^2}}\Big[2(n+1)(1+a^2)^2\, c_1\,r^3\nn\\ 
&&- (1+a^2)\big[[(2n\,(1+a^2)+(5-a^2))\,\rms+3(1+a^2)\,\rps]\,c_1-
 2\sqrt{2n}\,\,Q\,(1+a^2)\, c_2\big]\,r^2 \nn\\
&&+\big[\big(3\rms +7(1+a^2)\,\rps\big)\,\rms\, c_1 -
   2\sqrt{2n}\,\,Q\,(1+a^2)^2\, c_2\,\rms \big]\,r - 
   (4+a^2)\, c_1\,\rms^2\,\rps\Big]\,.\label{gaxial}
\eea 

  Up until this point, the calculations have proceeded rather analogously to
those that we performed when analysing the polar perturbation sector
in the EMD theory (and therefore also in the polar sector in our
present EMDA case).  However, we now encounter obstacles to proceeding 
further, for generic values of the coupling constants $a$, $b$ and $c$ 
in the theory.  Substituting the solutions for $f_i$ in eqns (\ref{tf123}) 
and the solution (\ref{gaxial}) for $V^-$ into the
remaining equations $e_3=0$ and $e_5=0$, gives the conditions
\crampest
\bea
&&\bigg(\frac{3-a^2}{1+a^2}\,\rms+3\,\rps  \bigg)\,c_1\,c_2 + 
2 \sqrt{2n}\,Q\Big(c_1{}^2- c_2{}^2\Big)
+2b\sqrt{2(n+1)}\,Q\,f_-^{\fft{a(2a+c)}{1+a^2}}\,c_1\, c_3=0\,,
\label{em3cond}\\
&&\sqrt{2n}\,\,Q\,r(r-\rms)\,c_2\,c_3 - 
b\sqrt{2(n+1)}\,Q\,r(r-\rms)\,f_-^\fft{a(2a+c)}{1+a^2}\,
c_1\, c_2-B(r)\, c_1\, c_3=0\,,\label{em5cond}
\eea
\uncramp
where $B(r)$ is quadratic in $r$ and independent of the constants $c_i$:
\crampest
\bea
B(r)&=& \fft1{2(1+a^2)^2}\,\Big[ 4(1+a^2)\,[(1+a^2)\rps + (1-a^2)\rms] \,
r^2 \nn\\
&& \quad + \rms\,[(a^2c^2-4)\, \rms - (1+a^2)(a c+10)\,\rps]\, r
  + (ac+2)(3+a^2-a c)\, \rms^2\, \rps\Big]\,.
\eea
\uncramp
Because of the occurrence generically of terms with incommensurate 
fractional powers 
of the metric function $f_-$,  equation (\ref{em3cond}) requires that 
we must impose the condition $c=-2a$.  This then implies that
\bea
B(r)= \fft{(r-\rms)}{1+a^2}\, \Big[
   2 [(1+a^2)\,\rps + (1-a^2)\,\rms]\,r -3(1-a^2)\,\rms\,\rps \Big]\,,
\eea
and therefore to be able to solve eqn (\ref{em5cond}) for all $r$
we must have $a=1$ (since, without loss of generality, we
may assume $a$ is non-negative).
Thus we can only solve the full system 
of equations $e_j=0$ if the constants $a$ and $c$ in the Lagrangian 
(\ref{emdalag}) are restricted to take the values 
\bea
a=1\,,\qquad c=-2\,.\label{accon2}
\eea
Accordingly, we shall assume from now on that eqns (\ref{accon2})
are imposed.

   With these assumptions, the equations $e_3=0$ and $e_5=0$
now imply the algebraic conditions
\bea
(\rms+3\,\rps)\, c_1\, c_2+2 \sqrt{2n}\,Q\big(c_1^2-c_2^2\big)
 +2b\sqrt{2(n+1)}\,Q\, c_1\, c_3=0\,,\label{em3condc2a1}\\
\Big(\sqrt{2n}\, c_3-b \sqrt{2(n+1)}\, c_1\Big)Q\,c_2-
  2\,\rps\, c_1\, c_3=0\,.\label{em5condc2a1}
\eea
Introducing a new dimensionless parameter $p$ in place of $c_1$ by defining
\bea
c_1=-\fft{p\, c_2\,\rms}{4\sqrt{2n}\,Q}\,,\label{cm1top}
\eea
eqn (\ref{em5condc2a1}) then implies
\bea
c_3=-\fft{b\sqrt{2(n+1)}\,Q\, c_2\,p}{2(p+2n)\rps}\,,\label{cm3top}
\eea
and so eqn (\ref{em3condc2a1}) implies that $p$ must satisfy the cubic
equation 
\bea
\tilde\Lambda(p) \equiv p^3\,\rms + p^2\,\big[2n\,(1+b^2)\,\rms - 
2(1-b^2)\,\rms-6\,\rps\big]- 4n\,p\,(\rms+7\,\rps)-32n^2\,\rps=0\,.
\label{pcubic}
\eea

   The final conclusion, then, is that we have three linearly independent
diagonalised axial eigenfunctions 
\bea
Z^-_{(p)} = c_1 \, \cG_{12}+  c_2\,B_{\u0\u3}+ c_3\,\chi\label{Zp}
\eea
obeying the second-order wave equations
\bea
(\del_{r_*}^2+\omega^2)Z^-_{(p)}  - V^-_{(p)} \,Z^-_{(p)} = 0\,,
\eea
where $c_1$ and $c_3$ are given in terms of $c_2$ by
eqns (\ref{cm1top}) and (\ref{cm3top}) and $p$ is any of the three roots of
eqn (\ref{pcubic}).  With the requirement that $r_+\ge r_-$ in eqn 
(\ref{pcubic}), $p$ is then restricted to lie in three specific, real ranges, 
as established below, and in Section \ref{p-ranges}.  
The corresponding potentials 
$V^-_{(p)}$ are given by substituting eqns (\ref{cm1top}) and (\ref{cm3top}) 
into eqn (\ref{gaxial}) with $a=1$; that is,
\bea
V^-_{(p)} &=&\fft{(r-\rps)\,\Big[8(n+1)\, r^3 
     -4\big( 2(n+1)\, \rms+ 3\rps\big)\, r^2
    + (3\rms+14\rps)\, \rms\, r - 5\rms^2\, \rps\Big]}{4 (r-\rms)^2\, r^4}\nn\\
  && -\fft{8n(r-\rps)\, \rps}{p\, (r-\rms)\, r^3} \,.\label{Vmp}
\eea

   It is straightforward to see, as we shall now show, that for all
valid values of the parameters, the three
roots $p_i$ of the cubic polynomial (\ref{pcubic}) are all real, and 
furthermore that two are negative and the third is positive:
The discriminant of cubic (\ref{pcubic}) is given by
\bea
D_p&=&64n^2\,\Big[16 (n+1)^3 \rms^3\, \rps\, b^6 +
  (n+1)^2 \,\big(\rms^2 + (48n-34)\rms\, \rps -95\rps^2\big)\,\rms^2\,  b^4 
\nn\\
&& + 2(n+1)[(n-1) \, \rms^3 +(24n^2+2n+7)\, \rms^2\, \rps +
       (157 n +53)\, \rms\, \rps^2 + 69 \rps^3]\, \rms\, b^2\nn\\
&& + [(n+1)\, \rms-\rps]^2 [\rms^2 +2(8n+3)\,\rms\,\rps + 9\rps^2]\Big]\,.
\label{disc}
\eea
This can be written as
\bea
D_p &=& 64n^2\, \Big[ 16(n+1)\, b^2\, \rms\, \rps\, 
[3\rps - (n+1)\, b^2\, \rms]^2 +
\rps^2\, [3\rps -(n+1)\, (3+b^2)\,\rms]^2 \nn\\
  && \quad\quad\quad +4(n+1)\,(77n+25)\,  b^2\, \rms^2\, \rps^2 
  +\rms\, \big[\rms +2(8n+3)\, \rps\big]\, [\rps -(n+1)\, \rms]^2 \nn\\
&& \qquad\quad+
2 (n+1)\, b^2\, \rms^3\, \big[(n-1)\rms + (24n^2+ 2n +7)\,\rps\big]\nn\\
&& \quad\quad\quad+ (n+1)^2\, b^4\, \rms^3\, 
\big[\rms + (48n-34)\rps\big]\Big]\,,
\eea
which is a sum of manifestly positive terms for all $b^2$, 
since we are concerned only with modes with $\ell\ge2$ and
therefore $n\ge 2$.  This establishes that the three roots $p_i$
are all real.

    From the cubic (\ref{pcubic}) it follows that the roots $p_i$
obey
\bea
p_1\, p_2\, p_3= \fft{32 n^2\, \rps}{\rms}>0\qquad \hbox{and}\qquad 
p_1\, p_2 + p_2\, p_3 + p_1\, p_3 = -\fft{4n\,(\rms+7\rps)}{\rms} <0\,.
\label{prodprodsum}
\eea
Ordering the roots as $p_1\le p_2\le p_3$, the first equation here implies
that either $p_i>0$ for all $i$, or else $p_1<0$ and $p_2<0$ with $p_3>0$.
The second equation in (\ref{prodprodsum}) implies that we cannot
have all $p_i>0$, and therefore it must be that
\bea
p_1\le p_2 < 0\,,\qquad p_3>0\,.
\eea

   Having successfully separated and diagonalised the equations
describing the linearised axial fluctuations around the static black hole
solutions in the $a=1$, $c=-2$ class of EMDA theories, we are in a 
position to study the question of mode stability.  This would be
established, by the energy integral method we used previously \cite{porowh}, 
if the potentials $V^-_{(p)}$ could be shown to be non-negative
everywhere outside the horizon at $r=\rps$.  Before addressing this
question for general values of the remaining parameter $b$ in the
EMDA theories (see eqn (\ref{emdalag})), we shall first consider the
special case where $b^2=1$.  This is of particular interest because 
then, the theory is actually closely related to 
the bosonic sector of an ${\cal N}=2$
supergravity.  As we shall now see, not only are the axial potentials
manifestly non-negative outside the horizon in this case but also,
we can establish a remarkable one-to-one mapping between the 
potentials, and also the modes, in the axial and the polar sectors.
This generalises earlier results obtained by Chandrasekhar for the
axial and polar modes in the perturbations of Schwarzschild \cite{chandra4} 
and Reissner-Nordstr\"om \cite{chandra3}.

\section{The Supergravity Case: $a=1$, $c=-2$, $b^2=1$}

\subsection{Relation between $V^-$ and $V^+$}

   If we assume not only $a=1$ and $c=-2$, as in eqn (\ref{accon2}), but
in addition
\bea
b^2=1\,,\label{bcon}
\eea
then the cubic equation (\ref{pcubic}) for $p$ becomes
\bea
\tilde\Lambda_1(p) \equiv p^3\,\rms + 
p^2\,\big[4n\,\rms-6\,\rps\big]- 4n\,p\,(\rms+7\,\rps)-32n^2\,\rps=0\,.
\label{pcubic1}
\eea
This is identical, with $p$ replaced by $q$, to the cubic equation 
that we found in \cite{porowh} characterising the diagonalised 
modes and potentials in the polar sector, after specialising that
result to $a=1$.  (See eqn (5.18) in \cite{porowh}.)  

  This raises the possibility that we might be able to establish a
one-to-one mapping between the potentials, and between the modes, 
of the axial sector and the polar sector in the EMDA theory.  
This would be a striking
extension of the analogous relation between the axial and polar
sectors of the perturbations of the Reissner-Nordstr\"om black holes,
which was established in \cite{chandra3} and elaborated upon in 
\cite{chandra}.

   The relation that was found in \cite{chandra3} between the
potentials in the axial and polar sectors for the Reissner-Nordstr\"om 
perturbations took the form
\bea
V^\pm = \pm \del_{r_*}\, f + \alpha\, f^2 + 2 \beta\,f \,,\label{Vpmexp}
\eea
where $f$ is related to a superpotential $W=f +\beta$,  
and $\alpha$ and $\beta$ are 
certain constants.   In our present case, we shall look for 
precisely such a relation (\ref{Vpmexp}) also.  Note that from 
eqn (\ref{Vpmexp}) it follows that
\bea
V^+ - V^- = 2 \del_{r_*}\, f\,.\label{feqn}
\eea
  
   The expression in \cite{porowh} for $V^+$ in the EMD case is quite
complicated, even under the specialisation $a=1$ that we now require,
and we shall not reproduce it here.  (It is given by the function $g$ in
eqn (5.14) of \cite{porowh}, and associated equations.) 

   Rather than 
viewing the potentials $V^\pm$ as being dependent on the black hole
parameters $\rms$ and $\rps$, with $p$ as being the roots of
the cubic equation (\ref{pcubic1}), it is easier to view $\rms$ and $p$
as being the two adjustable parameters, with $\rps$ then
being obtained by solving eqn (\ref{pcubic1}) to give
\bea
\rps= \fft{p\, \big[p^2+ 4n\, p -4n\big]\, \rms}{2 (2n+p)\, (8n+3p)}\,.
\label{rpsol}
\eea
Using this, and inserting the expressions for $V_{(p)}^\pm$ 
into eqn (\ref{feqn}), we find that
\bea
f_{(p)} = \fft{\rms\, (r-\rps)\, [(4n+2p)\, r -p\, \rms]}{2 r^2\, (r-\rms)\,
    (4nr + p\, \rms)}\,,\label{fsol}
\eea
It can now be verified that there do indeed exist constants $\alpha$
and $\beta$ such that eqns (\ref{Vpmexp}) both hold, with
\bea
\alpha_{(p)}=1\,, \qquad 
\beta_{(p)}=\frac{4n\,(n+1)}{(2n+p)\,\rms}\,,\label{albe}
\eea
and hence
\bea
V_{(p)}^\pm = \pm \del_{r*} f_{(p)} +  f_{(p)}^2 
   + \fft{8n\, (n+1)}{(2n+p)\,\rms}\, f_{(p)}\,.
\label{Vpmexp2}
\eea

\subsection{Relation between the $Z^+$ and $Z^-$ eigenfunctions}

  The radial equations in the axial and polar sectors obey
\bea
\Big[\del_{r*}^2 + \omega^2 -V^\pm\Big]\, Z^\pm =0\,, 
\label{Zeqns}
\eea
for each of the three roots $p$, 
with $-$ corresponding to the axial sector and $+$ to the polar sector.
In view of eqn (\ref{Vpmexp2}) it is evident that eqns (\ref{Zeqns}) can be
written as
\bea
D_-\, D_+\, Z^- + (\omega^2 +\beta^2)\, Z^- &=&0\,,\label{axialeqn}\\
D_+\, D_-\, Z^+ + (\omega^2 +\beta^2)\, Z^+ &=&0\,,\label{polareqn}
\eea
where we have defined
\bea
D_\pm = \del_{r*} \pm(f + \beta)\,,
\eea
and the constant $\beta$ is given in eqns (\ref{albe}).

  Acting with $D_+$ on the axial equation (\ref{axialeqn}) it can be seen 
that $D_+\, Z^-$ obeys the polar equation (\ref{polareqn}).  Similarly,
acting with $D_-$ on the polar equation (\ref{polareqn}) it can be seen 
that $D_-\, Z^+$ obeys the axial equation (\ref{axialeqn}).  In fact, we can
choose normalisations of the solutions so that the relations between
$Z^-$ and $Z^+$ take the form
\bea
D_-\, Z^+ = (\im \omega-\beta)\, Z^-\,,\qquad
D_+\, Z^- = (\im \omega+\beta)\, Z^+\,.
\eea
Thus we have an explicit one-to-one mapping between each mode of the
radial wave equation for the axial modes $Z^-$ and the polar modes
$Z^+$.  This generalises the analogous relation between the axial and
polar modes in the case of the perturbations of the Reissner-Nordstr\"om
black hole, which was established by Chandrasekhar in \cite{chandra3}.

\subsection{Algebraically Special modes}

   It was observed previously, for example in \cite{couchnewman,chandra6}, 
that there can exist certain exceptional solutions of the equations 
for linearised perturbations of black hole metrics that have the 
feature of being purely ingoing or purely outgoing.  These solutions,
which are not relevant when considering putative unstable modes obeying
physically-motivated boundary conditions on the horizon and at infinity,
have nonetheless been the subject of some investigations.  In the context
of perturbations of Schwarzschild or Kerr,
these modes were referred to as ``Algebraically Special'' perturbations, 
since they
preserved the algebraically-special form of the Weyl tensor.
In \cite{chandra6}, Chandrasekhar extended the notion
of Algebraically
Special to Reissner-Nordstr\"om, even though gravitational and
electromagnetic perturbations are inextricably coupled there.  Here, for
convenience, we do the same, and use the term ``Algebraically Special'' to
describe the similar, special generalisations that we find here of the
Schwarzschild and Reissner-Nordstr\"om modes.

    We observe from eqns (\ref{axialeqn}) and (\ref{polareqn}) 
that one can obtain such algebraically special solutions
for $Z^-$ and $Z^+$ having a very simple form, in the case where 
the frequency $\omega$ takes the purely imaginary values
\bea
\omega=\pm \im \beta\,,\label{specom}
\eea
and where $Z^+$ or $Z^-$ obey the first-order equations 
\bea
 D_+\, Z^-=0\,, \qquad D_-\, Z^+=0\,,\label{firstorder}
\eea
respectively.  Thus the algebraically special solutions are given by
\bea
Z^\pm = \exp\Big[\pm \int^r \big(f(r') + \beta\big)\,
         e^{-2\bnu(r')}\, dr'\Big]\,.\label{specsol2}
\eea
Substituting in the expressions for $f_{(p)}$ and $\beta_{(p)}$ from eqns 
(\ref{fsol}) and (\ref{albe}), we find 
\bea
Z_{(p)}^+ &=& e^{\beta_{(p)} r_*}\,\, \Big(1-\fft{\rms}{r}\Big)^{\ft12}\,
\Big(4n + \fft{p\, \rms}{r}\Big)^{- 1}\,,\nn\\
Z_{(p)}^- &=& e^{-\beta_{(p)} r_*}\,\, \Big(1-\fft{\rms}{r}\Big)^{-\ft12}\,
\Big(4n + \fft{p\, \rms}{r}\Big)\,,\label{Zpmodes0}
\eea
where $r_*=\int^r e^{-2\bnu(r')}\, dr'= 
 r+\rps\, \log\Big(\dfft{r}{\rps}-1\Big)$ is the tortoise coordinate.
(Note that these algebraically special polar and axial solutions
are inversely related, with $Z_{(p)}^+\, Z_{(p)}^- =1$.)
Including the time dependence $e^{-\im\omega t}$, with $\omega=\pm\im \beta$,
these special explicit solutions have the forms
\bea
e^{-\im\omega t}\, Z_{(p)}^+ &=& 
 e^{\beta_{(p)}\, (r_*\pm \,t)}\,\, \Big(1-\fft{\rms}{r}\Big)^{\ft12}\,
\Big(4n + \fft{p\, \rms}{r}\Big)^{- 1}\,,\nn\\
e^{-\im\omega t}\, Z_{(p)}^- &=& 
  e^{-\beta_{(p)} \,(r_* \mp \,t)}\,\, \Big(1-\fft{\rms}{r}\Big)^{-\ft12}\,
\Big(4n + \fft{p\, \rms}{r}\Big)\,,\label{Zpmodes}
\eea
Note that although these are of interest as explicit exact solutions of the
linearised equations, they do not obey physically interesting boundary
conditions.  This is because 
if they are ingoing on the future horizon they are
also ingoing at past null infinity $\scri^-$, and conversely if outgoing on
future null infinity $\scri^+$ they are also outgoing on the
past horizon.

  As well as the pair of algebraically special solutions obtained above,
one can also, of course, construct in each case 
the corresponding ``second solution'' by the standard Wronskian method.
Thus one also has the explicit (up to quadrature) second solutions of
the form
\bea
\tilde Z_{(p)}^+(r) = Z^+_{(p)}(r)\, 
\int^r\fft{e^{-2\bnu(r')}}{\big(Z^+_{(p)}(r')\big)^2}\, dr'\,,\qquad
\tilde Z_{(p)}^-(r) = Z^-_{(p)}(r)\, 
\int^r\fft{e^{-2\bnu(r')}}{\big(Z^-_{(p)}(r')\big)^2}\, dr'\,,
\label{tildedZ}
\eea
where $Z_{(p)}^+$ and $Z_{(p)}^-$ are given in eqns (\ref{Zpmodes0}). 
The integrals in eqns (\ref{tildedZ}) are rather involved to evaluate
explicitly.  We just remark here that these second
solutions $\tilde Z_{(p)}^+$ and
$\tilde Z_{(p)}^-$ do not have the feature of being purely ingoing or
purely outgoing, unlike the solutions (\ref{Zpmodes0}).

\subsection{Positivity of the $V^-$ potentials}

   The positivity of the axial potentials $V^-$ outside the horizon
in the supergravity case $a=1$, $c=-2$, $b^2=1$ may be established using
the same techniques that we employed for the polar potentials $V^+$
in the general EMD case in \cite{porowh}.  

    We already showed in section \ref{diagsec} 
that for valid values of the parameters 
the three roots $p$ of the cubic equation (\ref{pcubic1}) 
are real, and furthermore two are negative and one positive.  
Thus when we instead choose to
solve (\ref{pcubic1}) for $\rps$, as in eqn (\ref{rpsol}), there will
be three disjoint real ranges for $p$, two negative and one positive, 
that correspond to the situations 
of interest to us where $\rps>\rms$. Writing $n$ in terms of $\ell$, as in eqn
(\ref{ndef}), the solution for $\rps$ in eqn (\ref{rpsol}) implies
that $(\rps-\rms)$ factorises over the rationals,
\bea
\rps - \rms= \fft{[p+2(\ell-1)]\,[p-2(\ell+2)]\, 
   [p + 2 (\ell+2)(\ell-1)]\, \rms}{2[p+(\ell+2)(\ell-1)]\,
    [3p+ 4(\ell+2)(\ell-1)]}\,.
\eea
Thus we can easily identify the three segments in the ranges of $p$ for which
$(\rps-\rms)> 0$, namely
\bea
{\bf(1)}:&& p> 2(\ell+2)\,,\nn\\
{\bf(2)}:&& -(\ell+2)(\ell-1)< p < -2(\ell-1)\,,\nn\\
{\bf(3)}:&& -2(\ell+2)(\ell-1) < p < -\ft43 (\ell+2)(\ell-1)\,.
\label{cases}
\eea

   Generically, we need to consider modes with $\ell\ge 2$.  We can then
parameterise $\ell$ and $r$ (outside the horizon) by
\bea
\ell= 2+y\,,\qquad  r= \rps\,(1+z)\,,
\eea
where $y\ge0$ and $z\ge0$.  For the three cases in eqns (\ref{cases})
we can parameterise the valid ranges of $p$ in terms of $x\ge0$,
with
\bea
\hbox{\bf Case (1)}: && p=2(\ell+2)\, (1+x)\,,\nn\\
\hbox{\bf Case (2)}: && p= -(\ell+2)(\ell-1)+ \fft{\ell(\ell-1)}{1+x}\,,\nn\\
\hbox{\bf Case (3)}: && p= -2(\ell+2)(\ell-1) + 
                    \fft{2(\ell+2)(\ell-1)}{3(1+x)}\,.
\eea
For each of the three cases, we can plug these parameterisations into the
expression (\ref{Vmp}) for the potential $V_{(p)}^-$.  
We find that in each case the
potential $V^-_{(p)}$ then takes the factorised form
\bea
V^- = R\, P\,,
\eea
where for each of the three cases $R$ is a simple manifestly-positive 
rational function of
$x$, $y$ and $z$, and $P$ is a somewhat complicated multinomial
in $x$, $y$ and $z$. In fact in case (1) $P$ has 177 terms; 
in case (2) $P$ has 207 terms; and in case (3)
$P$ has 169 terms.  Remarkably, 
the coefficient of each term in each of the three multinomials 
$P$ is positive, thus
proving that $V^-$ is positive outside the horizon for each of the three
cases in (\ref{cases}).

   It was already established in \cite{porowh} that the polar potential
$V^+$ was positive outside the horizon for the case of static black holes
in the whole class of EMD theories with arbitrary values of $a$.  Thus in
particular, $V^+$ is positive outside the horizon for the case $a=1$, 
which corresponds to the potential in the polar sector in our 
present analysis of the perturbations around static black holes in
the $a=1$, $c=-2$, $b^2=1$ EMDA theory. Thus, combined with our
new analysis here of the axial sector in this theory, we have obtained
a complete demonstration of the linearised mode stability of the
static black holes in this EMDA theory.  

Some illustrative plots showing examples of
the three 
axial and three polar potentials, for the case of the $\ell=2$ modes, 
can be found in Figure 1 in appendix \ref{Plots}.  In particular, it
can be seen that for $b=1$, all the potentials are indeed positive outside the
horizon, consistent with the positivity proof we gave above.

\subsection{Relation to ${\cal N}=2$ supergravity \label{STUsec}}

   STU supergravity is ${\cal N}=2$ supergravity coupled
to three vector multiplets \cite{dulira}. 
Thus its bosonic sector comprises the metric,
four $U(1)$ Maxwell gauge fields, three dilatons and three axions. There
exists a supersymmetric consistent truncation in which two dilatons and two
axions are set to zero and the four Maxwell fields are set pairwise equal,
together with corresponding truncations in the fermionic sector.  
(The truncation is discussed in a number of papers, including, for example,
\cite{chowcomp,cvposa}.)  The
bosonic sector of this truncated theory is described by the Lagrangian
\bea
{\cal L}&=& \sqrt{-g}\, \Big[ R -\ft12 (\del\phi)^2 - e^{\phi}\, F^2
-\ft12 e^{-2\phi}\,(\del\chi)^2 +
   \,\chi\,\widetilde{F}^{\mu\nu}F_{\mu\nu} \nn\\
&& \qquad\quad -\fft{e^{-2\phi}}{1+\chi^2\, e^{2\phi}} \,\big(e^\phi\, H^2 +
     \chi\, \widetilde{H}^{\mu\nu}H_{\mu\nu}  \big)  \Big]\,,
\label{truncSTUlag}
\eea
where $H_{\mu\nu}$ is the field strength of the second Maxwell field.
Note that the first line in (\ref{truncSTUlag}) is exactly the same as
our EMDA Lagrangian (\ref{emdalag}) when $a=1$, $c=-2$ and $b=1$.
It can be seen that the Gibbons-Maeda static black hole solutions
that we are taking to form the backgrounds in the EMDA theories
will also be solutions of the theory (\ref{truncSTUlag}), in which
we take the additional Maxwell field to vanish, $H_{\mu\nu}=0$.  Furthermore,
the linearised perturbations around the static Gibbons-Maeda
black holes within this supergravity theory will comprise precisely the
modes we have already obtained in the EMDA theory with
$a=1$, $c=-2$, $b^2=1$, together with a rather simple and totally
decoupled system of modes for the $H_{\mu\nu}$ field, obeying
the equation $\bar\nabla^\mu (e^{-\bar\phi}\, \delta H_{\mu\nu})=0$.

\subsection{Inclusion of the $H_{\mu\nu}$ field of the supergravity 
              theory \label{Hperturbations}}

  As discussed in section \ref{STUsec}, we can also view the Gibbons-Maeda
black hole for $a=1$ as being a solution of the ${\cal N}=2$ supergravity
whose bosonic sector is described by the Lagrangian (\ref{truncSTUlag}).
As we mentioned, the complete set of linearised bosonic perturbations
within this theory will comprise exactly the perturbations within the 
EMDA theory as we have discussed already earlier in this section, together with
a decoupled system describing the perturbations of the $H_{\mu\nu}$ field;
these obey $\bar\nabla^\mu(e^{-\bar\phi}\, \delta H_{\mu\nu})=0$.  (As
usual, the overbars denote background quantities.)  These perturbations
will themselves divide into disjoint axial and polar sectors.  

   Because of
the fact that the $\delta H_{\mu\nu}$ perturbations are decoupled from
the other perturbations in the EMDA sector, it is in fact rather straightforward
to implement the separation of variables, and the construction 
of the second-order radial equations for each of the axial and polar
perturbations, by making appropriate truncations and modifications to
results we have already presented here and in \cite{porowh}.  For
example, in the axial sector we just take the Bianchi identities 
(\ref{rbianchi1}) and
(\ref{rbianchi2}), with the components $B_{\u a\u b}$  replaced by 
the corresponding components $C_{\u a\u b}$ associated with the 
$\delta H_{\u a\u b}$
perturbations, and then we take the Maxwell equation 
(\ref{rmax3}), with $B_{\u a\u b}$
replaced by $e^{-2\bar\phi}\, C_{\u a\u b}$ (to take account of
the opposite sign of the dilaton coupling in the exponential
prefactor in the kinetic term for $H_{\u a\u b}$).  
At the same time the $\cG_{01}$
and $\chi$ terms should be omitted (since $\delta H_{\u a\u b}$ 
does not couple to any
other perturbations).  In the polar sector, we can analogously read off the 
equations from eqns (4.16) -- (4.18) of \cite{porowh}, sending $B_{\u a\u b}
\rightarrow C_{\u a\u b}$ in the Bianchi identity and $B_{\u a\u b}\rightarrow 
e^{-2\bar\phi}\, C_{\u a\u b}$ in the Maxwell equations, and also with the
perturbations of the metric and the dilaton field omitted. In each of the
axial and polar sectors, appropriate additional $r$-dependent 
scaling redefinitions
are also then needed, in order to arrive at radial functions 
$Z^\pm_{\sst H}$ that obey standard $(\del_{r_*}^2 + \omega^2 -
   V^\pm_{\sst H})\,Z^\pm_{\sst H}=0$ wave equations.

   The upshot is that in the axial sector we can write
\bea
\delta H_{\u0\u3}(t,r,\theta)  &=& -\fft{\im\omega\, e^{-\bnu+\bar \phi}}{r}\, 
e^{-\im\omega t}\, Z_{\sst H}^-(r)\, \del_\theta P_\ell(\cos\theta)\,,
\eea
and in the polar sector
\bea
\delta H_{\u1\u2}(t,r,\theta) &=&  
-\fft{\im\omega\, e^{-\bnu +\bar\phi}}{r^2}\,
            e^{-\im\omega t}\, Z_{\sst H}^+(r)\, \del_\theta 
P_\ell(\cos\theta)\,,
\eea
with the other components of $\delta H_{ab}$ being expressed just in
terms of $Z^-_H(r)$ and $Z^+_H(r)$. Specifically, we find
\bea
\delta H_{\u1\u3} &=& e^{\bnu-\bmu_2}\, \del_r \big(r\,e^{-\bmu_2}\, 
Z^-_{\sst H}(r)\big)\, e^{-\im\omega\,t}\, \del_\theta P_\ell(\cos\theta)\,,\nn\\
\delta H_{\u2\u3} &=&-\fft{\ell(\ell+1)\, e^{-\bmu_2+\bar\phi}}{r}\, 
      Z^-_{\sst H}(r)\,  e^{-\im\omega\,t}\, P_\ell(\cos\theta)\,,
\eea
in the axial sector, and
\bea
\delta H_{\u1\u2} &=& -\fft{\im\omega\, e^{-\bnu+\bar\phi}}{r^2}\, 
Z^+_{\sst H}(r)\, e^{-\im\omega\, t}\, \del_\theta P_\ell(\cos\theta)\,,\nn\\
\delta H_{\u0\u1} &=& \fft{\ell(\ell+1)\, e^{-\bmu_2+\bar\phi}}{r^2}\,
   Z^+_{\sst H}(r)\, e^{-\im\omega\,t}\, P_\ell(\cos\theta)\,,
\eea
in the polar sector.

     The radial equations then turn out to be
\bea
(\del_{r_*}^2 +\omega^2 -V^\pm_{\sst H})\, Z^\pm_{\sst H}=0\,,
\eea
where the axial and polar potentials are given by
\bea
V_{\sst H}^\pm = \fft{2(n+1)\, e^{2\bnu +\bar\phi}}{r^2} +
    e^{4\bnu}\, \Big[\ft14 {\bar{\phi'}}^2 \pm \big(\bnu'\, \bar\phi' 
                 +\ft12 \bar\phi''\big)\Big]\,,\label{VHpm0}
\eea
Thus we can find a function $f_{\sst H}$ such that
\bea
V^\pm_{\sst H} = \pm \del_{r_*} f_{\sst H} + f_{\sst H}^2 +
  2\beta_{\sst H}\, f_{\sst H}\,, \label{VHpm1}
\eea
with
\bea
f_{\sst H} = \ft12 e^{2\bnu}\, \bar\phi' \,,\qquad
\beta_{\sst H} = \fft{2(n+1)\,
e^{\bar\phi}}{r^2\, \bar\phi'} \,.\label{fHbeta}
\eea

  Explicitly, the two potentials in eqn (\ref{VHpm0}) are given by
\bea
V^-_{\sst H} &=& \fft{(r-\rps)[8(n+1)\, r^3 - 
  4(2n+3)\, \rms\, r^2 + 3(\rms+2\rps)\, \rms\, r 
           -5 \rms^2\, \rps]}{4 r^4\, (r-\rms)^2}\,,\\
&&\nn\\
V^+_{\sst H} &=& \fft{(r-\rps)[8(n+1)\, r^3 -
  4(2n+1)\, \rms\, r^2 - (\rms+6\rps)\, \rms\, r
           +3 \rms^2\, \rps]}{4 r^4\, (r-\rms)^2}\,,
\eea
and, as can be seen from eqns (\ref{bhsol}), the function $f_{\sst H}$ and 
the constant $\beta_{\sst H}$ in eqns (\ref{fHbeta}) are given by
\bea
f_{\sst H}= -\fft{\rms\, (r-\rps)}{2 r^2\, (r-\rms)}\,,\qquad
   \beta_{\sst H} = -\fft{2(n+1)}{\rms}\,.\label{VHpmexp}
\eea
It can easily be seen, by writing $r=(1+z)\, \rps$, \ $\rps=(1+w)\, \rms$
and $n=2+y$, that $V^+_{\sst H}$ and $V^-_{\sst H}$ are both non-negative
everywhere outside the horizon.

   In the case of these axial and polar modes of
the $H_{\mu\nu}$ field, the explicit algebraically special solutions that we
obtained in eqns (\ref{firstorder}) and (\ref{specsol2}) for the axial and
polar modes in the EMDA sector are extremely simple, and we find
\bea
\omega=\pm\im\beta_{\sst H}= \mp\fft{2\im (n+1)}{\rms}\,,
\eea
\bea Z_{\sst H}^+ = e^{\beta_{\sst H}\, r_*}\, 
           \Big(1-\fft{\rms}{r}\Big)^{-\ft12}\,,
\qquad
Z_{\sst H}^- = e^{-\beta_{\sst H}\, r_*}\,
           \Big(1-\fft{\rms}{r}\Big)^{\ft12}\,,
\eea
where $r_*= r + \rps\,\log\Big(\dfft{r}{\rps}-1\Big)$ is the tortoise
coordinate.  For the two sign choices for $\omega$ in these special
solutions, the $t$ and $r$ dependences for these waves have the form
\bea
e^{-\im\omega t}\, Z_{\sst H}^+ =
   e^{\beta_{\sst H} \, (r_* \pm\, t)}\,
           \Big(1-\fft{\rms}{r}\Big)^{-\ft12}
\eea
and
\bea
e^{-\im\omega t}\, Z_{\sst H}^- =
   e^{-\beta_{\sst H}\, (r_* \mp \, t)}\,
           \Big(1-\fft{\rms}{r}\Big)^{\ft12}\,.
\eea
For the same reason as we already discussed for the triplets of
explicit modes $Z_{(p)}^\pm$ obtained in eqns (\ref{Zpmodes0}),
these exact solutions do not obey physically interesting 
boundary conditions.  The second solutions of the analogous forms
to those given in eqns (\ref{tildedZ}) can also be constructed
here.  In fact it happens that in the polar sector, the
second solution $\tilde Z_{\sst H}^+$ turns out to have a relatively simple
explicit form, namely
\bea
\tilde Z_{\sst H}^+ = \fft{\Big(2\beta_{\sst H}\,\rms\, 
   e^{\beta_{\sst H} (r-2\rps)}\,
   {\rm Ei}[1+2\beta_{\sst H}\, \rps,
  2\beta_{\sst H}\,(r-\rps)] -e^{-\beta_{\sst H}\, r}\Big)\, \sqrt{r}}{
2\beta_{\sst H}\, \sqrt{r-\rms}\, (r-\rps)^{\beta_{\sst H}\, \rps}}\,,
\eea
where ${\rm Ei}[\alpha,z]= \int_1^\infty e^{-z t}\, t^{-\alpha}\, dt$ is
the exponential integral function.

 To summarise, we have shown that if we view the $a=1$ Gibbons-Maeda 
charged static black holes as solutions in the ${\cal N}=2$ supergravity
theory whose bosonic sector is described by the Lagrangian 
(\ref{truncSTUlag}), then the complete set of linearised perturbations
are described by the radial functions $Z^\pm_i$, for $i=1$, 2 and 3, together
with the radial functions $Z^\pm_{\sst H}$.  For each of these,
the associated potentials $(V^+_1, V^+_2, V^+_3, V^+_{\sst H})$ in the polar
sector are related to the corresponding potentials
$(V^-_1, V^-_2, V^-_3, V^-_{\sst H})$ in the axial sector, as given 
in eqns (\ref{Vpmexp2}) and (\ref{VHpm1}).   All the potentials
are non-negative outside the outer horizon, thus demonstrating linearised
mode stability for the $a=1$ Gibbons-Maeda black holes as solutions of
${\cal N}=2$ supergravity.

\section{The $a=1$, $c=-2$ EMDA Theories With $b^2$ Arbitrary}

  When $b^2\ne 1$, the axial sector of the $a=1$, $c=-2$ EMDA theories
is no longer related to the polar sector in the way we saw in the $b^2=1$
case in the previous section.  It is still of interest, however, to
investigate the properties of the potential $V^-$ in the axial
radial wave equation.  

\subsection{Allowed ranges for $p$ \label{p-ranges}}

   When $b^2$ is arbitrary, we can again view the cubic equation
$\tilde\Lambda(p)$ in eqn (\ref{pcubic}) as a linear equation for
$\rps$, with $p$ rather than $\rps$ being taken to be a parameter of
the solutions.   As in the $b^2=1$ case, there are generally three 
allowable ranges 
for $p$, now dependent on $b$, which we find by examining the 
expression for $\rps-\rms$, namely
\bea\label{delta_rp-m}
\rps - \rms =\frac{\rms\, \big(p^3+[2n(1+b^2)-2(1-b^2)-6]
p^2-32np-32n^2\big)}{2(p+2n)(3p+8n)}\,,\label{rprm}
\eea
and requiring $(\rps-\rms)>0$.  

   There are two special values for $b^2$ such that the numerator 
of (\ref{rprm})
has a factor which is also in the denominator, namely when 
$b^2=0$ and when $b^2=\dfft{4n+3}{12(n+1)}$.  These correspond to
cases where the cubic (\ref{pcubic}) factorises, with a linear factor
that is independent of $\rps$ and $\rms$:
\bea
b^2=0:&& \tilde\Lambda(p)= (p+2n)\,[\rms\, p^2 -2(\rms+3\rps)\, p-16n\, \rps]
\,,\nn\\
b^2=\fft{4n+3}{12(n+1)}: && \tilde\Lambda(p)= 
\fft16 (3p+8n)\, [2 \rms\, p^2 -3(\rms+4\rps)\,p -24n\, \rps]\,.
\eea
In each of these cases there is therefore a discrete root for $p$ that is
independent of $\rps$ and $\rms$, and the two other roots 
lie at points within ranges that depend upon the
ratio $\dfft{\rps}{\rms}$ (which must exceed 1). Thus, when $b^2=0$ we have
\bea
\hbox{\bf Fixed}: && p=-2n\,,\\
\hbox{\bf Range (i)}:  && -\frac{8n}{3}<p<4(1-\sqrt{n+1})<0\,, \\
\hbox{\bf Range (ii)}: && 0<4(1+\sqrt{n+1})<p<\infty\,,
\eea
and when instead $b^2=\dfft{4n+3}{12(n+1)}$ we have
\bea
\hbox{\bf Fixed}: && p=-\fft{8n}{3}\,,\\
\hbox{\bf Range (i)}: &&
     -2n<p<\frac{15}{4}(1-\sqrt{1+\frac{64n}{75}})<0\,,\\
\hbox{\bf Range (ii)}: &&
 0<\frac{15}{4}(1+\sqrt{1+\frac{64n}{75}})<p<\infty\,.
\eea

   For all other values of $b^2>0$, there are three valid ranges for $p$, 
and these divide into two classes, one for $0<b^2<\dfft{4n+3}{12(n+1)}$, 
and the other for $\dfft{4n+3}{12(n+1)}<b^2<\infty$.  
In the first case, the numerator in eq (\ref{delta_rp-m}) has a zero, 
$p_z=p_a$, between $-\dfft{8n}{3}$ and $-2n$, and an allowable range for 
$p$ is $-\dfft{8n}{3}<p<p_a<-2n$;  in the second case, the numerator has a 
zero, $p_z=p_b$, between $-\infty$ and $-\dfft{8n}{3}$, and an allowable 
range for $p$ is $-\infty<p_b < p < -\dfft{8n}{3}$.  Using the location of 
each zero, we can solve $\rps-\rms=0$ in eqn (\ref{rprm}) 
for $b^2$ in terms of that zero 
($p_z=p_a$ or $p_z=p_b$):
\bea
b^2=\frac{(p_z+2n)\, [2(\ell+2)-p_z]\, [p_z+2(\ell-1)]}{2p_z^2(n+1)}>0 \quad 
\forall \, p_z<-2n,
\eea
and then factor the numerator of eq (\ref{delta_rp-m}) into 
$(p-p_z)(p-p_-)(p-p_+)$, where $p_{\pm}$ are solutions (see below) to 
the resulting quadratic equation.  The additional allowable ranges for 
$p$ are then $-2n<p<p_-<0$ and $0<p_+<p<+\infty$.  

While it would be most convenient to express our results in terms of $b^2$, the task of then solving the cubic for $p$ makes this impractical.  What we have found is that, as we change $b^2$ from $0$ to $\infty$, one root of the cubic moves from $p=-2n$ to $-\infty$.  If we choose a $p$ in this latter range, then we can simultaneously find the value of $b^2$ for which this is a root, and solve the remaining quadratic for the other two roots for $p$.  This is the approach we use, whereby we can obtain 
appropriate ranges for all roots $p$, and for $b^2$, that are parameterised 
in terms of a real variable $x$ lying in the range $0\!<\!x\!<\!\infty$:  

  For $b^2$ lying in the range $0<b^2 <\dfft{4n+3}{12(n+1)}$, it will
be parameterised as
\bea
b^2=\frac{4(4n+3)x^3+12(2n+1)x^2+9nx}{3(n+1)(3+4x)^2(1+x)}\,,\qquad
{\rm with}\quad 0<x<\infty\,,
\eea
and we find that the three valid ranges for $p$ are
\crampest
\bea
\hbox{\bf Range (i)}:&& \mkern-18mu
  -\frac{8n}{3}<p<p_a<-2n\quad{\rm where}\quad 
p_a=-\,{\frac {2n \left( 3+4\,x \right) }{3(1+x)}},\\
\hbox{\bf Range (ii)}: && \mkern-18mu
  -2n<p<p_{-,a}<0\quad{\rm where}\quad p_{-,a}=
 \frac{4\left(3(1+x)(3+5\,x)-\sqrt{P_a}\right)}{\left( 3+4\,x \right) ^{2}}
,\ \ \\
\hbox{\bf Range (iii)}:&&\mkern-18mu
  0<p_{+,a}<p<\infty\quad{\rm where}\quad 
p_{+,a}=\frac{4\left(3(1+x)(3+5\,x)+
   \sqrt{P_a}\right)}{\left( 3+4\,x \right)^{2}},\\  
 {\rm where}&& P_a=3n(1+x)(3+4x)^3+\left[3(1+x)(3+5x)\right]^2\,.
\eea
\uncramp

    Similarly, for $b^2$ lying in the range $\dfft{4n+3}{12(n+1)} <b^2<\infty$
it will be parameterised as 
\bea
b^2=\frac{(1+4x)(4nx^2+8nx+4n+12x+3)}{12(1+x)^2(n+1)}<\infty\,, 
\qquad {\rm with}\quad 0<x<\infty\,,
\eea
and we find that the three valid ranges for $p$ are
\bea
\hbox{\bf Range (i)}:&&
  -\infty<p_b<p<-\frac{8n}{3}\quad{\rm where}\quad p_b=-\frac{8n}{3}(1+x),\\
\hbox{\bf Range (ii)}:&&
  -2n<p<p_{-,b}<0\quad{\rm where}\quad 
p_{-,b}=\frac{3(5+8x)-\sqrt{P_b}}{4(1+x)^2},\\
\hbox{\bf Range (iii)}:&& 
0<p_{+,b}<p<\infty\quad{\rm where}\quad p_{+,b}=
  \frac{3(5+8x)+\sqrt{P_b}}{4(1+x)^2},\\ 
{\rm where}&& P_b=3n\left[4(1+x)\right]^3+\left[3(5+8x)\right]^2\,.
\eea
Note that the special case $b^2=1$ here 
corresponds to $x=\half$.

   It should also be noted that in all cases, we have that for any given
choice of the parameter values, there are two negative roots and one positive
root, in accordance with what we found in section \ref{diagsec} and
section \ref{p-ranges}.

\subsection{Conditions under which $V^-$ can be negative outside the
horizon}

   The three axial potentials, corresponding to the three roots $p_i$
of the cubic polynomial (\ref{pcubic}), are given by eqn (\ref{Vmp}).
Note that the parameter $b$ does not appear explicitly in the
expression (\ref{Vmp}); it enters only in the cubic (\ref{pcubic}) 
whose roots determine the three values $p_i$ for the parameter $p$.
Note also that the potential $V^-_{(p)}$ in eqn (\ref{Vmp}) is
of the form
\bea
V^-_{(p)} = -\fft{A}{p} + B\,,\label{VmAB}
\eea
where $A$ and $B$ are independent of $p$, and given by
\bea
A&=& \fft{8n(r-\rps)\, \rps}{(r-\rms)\, r^3}\,,\label{ABexp}\\
B&=& \fft{(r-\rps)\,\Big[8(n+1)\, r^3 
     -4\big( 2(n+1)\, \rms+ 3\rps\big)\, r^2
    + (3\rms+14\rps)\, \rms\, r - 5\rms^2\, \rps\Big]}{4 (r-\rms)^2\, r^4}\,.
\nn
\eea

   The function $A$ is manifestly positive outside the outer horizon, and
by writing
\bea
r=\rps \,(1+z)\,,\qquad \rps=\rms\, (1+w)
\eea
it follows that $B$ can be written as
\bea
B&=&  \fft{(r-\rps)\,\rms^2\, \rps}{4 r^4\, (r-\rms)^2}\,\Big[
8(n+1)(1+w)^2\, z^3 + 4(1+w)\,[4n+1 + 3(2n+1)\, w]\,z^2\nn\\
&&
\qquad + [8n+1 + 2(16n-1)\, w +24 n\, w^2]\, z +
2w\, [4n-1 + 2(2n-1)\, w]\Big]\,.
\eea
Since $n\ge2$ it can be seen that $B$ is also manifestly positive outside the
outer horizon.  It follows that if $p$ is negative then the
potential $V^-_{(p)}$ is definitely positive outside the horizon. In other
words, the axial potential $V^-_{(p)}(r)$ 
only has the possibility of being negative
outside the horizon in the case where the root $p$ is positive.  It may
be recalled that we showed in section \ref{diagsec} that the three roots
are always real, and that two of them are negative while the third is
positive. 

Numerical examples show that under certain circumstances the potential
can in fact be negative in some region outside the horizon.  For example,
if the parameters are taken to be such that
\bea
\rps=2 \,\rms \,,\qquad \ell=2\,,\qquad b=5\,,
\eea
then for the positive root, $p\approx 1.83155$, the potential is 
negative in the region $\rps< r< r_{\rm max}$ near to the horizon, 
where $r_{\rm max} 
\approx 1.925 \, \rps$. It is of interest, therefore, to 
investigate more extensively to establish when the potential can be negative
outside the outer horizon.

\subsection{Some properties of the potential}

  The potential $V^-_{(p)}(r)$ takes the form
\bea
V^-_{(p)} = \fft{(r-\rps)}{4 r^4\, (r-\rms)^2}\, U(r)\,,
\eea
where
\bea
U(r) &=& 8(n+1)\, r^3 -4 [(2(n+1) \,\rms + 3\rps] \,r^2 +
    (3\rms+14\rps)\, \rms\, r
- 5 \rms^2\, \rps \nn\\
&&
 - \fft{32 n (r-\rms)\, \rps\, r}{p}\,.\label{Udef}
\eea
As we have already observed, the potential outside the horizon cannot be
negative unless $p$ is positive.  The potential 
will be negative there if and only if
$U(r)$ is negative for some $r>\rps$.

   The function $U(r)$ is cubic in $r$, and therefore has three roots
$r_1$, $r_2$ and $r_3$.  These are either all real, or else one is real
with the other two being a complex-conjugate pair.  In the limits of
large negative and positive $r$ we have
\bea
U(r)\longrightarrow -\infty &&\qquad \hbox{when}\quad r\longrightarrow -\infty
\,,\nn\\
U(r)\longrightarrow +\infty && \qquad \hbox{when}\quad r\longrightarrow +\infty
\,.
\eea
We also have
\bea
U(0)= -5 \rps\, \rms^2\, <0\,,\qquad U(\rms)= -3(\rps-\rms)\, \rms^2 \,
<0\,,
\label{Uvals}
\eea
and
\bea
r_1\, r_2\, r_3 =\fft{5}{8(n+1)}\, \rps\, \rms^2 \, >0\,.\label{r1r2r3}
\eea
If the roots are all real, then with $r_1\le r_2\le r_3$ it must be
that $r_3$ at least is positive ($r_1$ and $r_2$ could be both positive
or both negative).  If there is only one real root, $r_3$, then $r_3$ must
again be positive.  We can now establish the following:
\bigskip

\noindent
{\bf (1) $U(r)$ can have at most one root outside the horizon at $\rps$}:
\medskip

   It follows from eqn (\ref{r1r2r3}) that in the case of three real
roots, at most one can lie outside the horizon at $\rps$:  Suppose
all three were outside $\rps$; in this case
\bea
\fft{r_1}{\rps}\, \fft{r_2}{\rps}\, \fft{r_3}{\rps} =
  \fft{5}{8(n+1)}\, \fft{\rms^2}{\rps^2}\,,
\eea
implying a contradiction since the left-hand side would be greater than 1
while the right-hand side would be less than 1.  Suppose instead 
two roots lay outside the outer horizon; in this case
\bea
\fft{r_1}{\rms}= \fft{5}{8(n+1)}\,\fft{\rps}{r_3}\, \fft{\rms}{r_2}\,.
\eea
The right-hand side would be less than 1, and hence $0<r_1<\rms$.  But
there cannot exist just one root between $0$ and $\rms$, since as seen 
in eqn (\ref{Uvals}), $U(r)$ is negative both at $r=0$ and $r=\rms$.  Thus
$U(r)$ can have at most one root outside the outer horizon.  

    Suppose, then, that there exists just one root, $r_3$, outside the
outer horizon.  Solving $U(r_3)=0$ for $p$ and substituting this back
into $U(r)$ gives
\bea
U(r)= \fft{(r-r_3)}{r_3\, (r_3-\rms)}\, P_2(r)\,,
\eea
where 
\bea
P_2(r) &=& 8(n+1)\,r_3\,(r_3-\rms)\, r^2 - \rms\, [8(n+1) \,r_3^2
 -(8n+5)\, r_3\, \rms + 2 r_3\, \rps - 5\rms\, \rps]\, r\nn\\
&&  +
  5(r_3-\rms)\,\rms^2\, \rps\,.
\eea
Letting $\rps=\rms\, (1+w)$, $r=\rps\, (1+z)$ and $n=2+y$ it can be seen
that for positive $y$, $z$ and $w$ the discriminant of the quadratic $P_2(r)$
is positive, and furthermore that the two (real) 
roots $r_1$ and $r_2$ must both 
be positive, with $0<r_1\le r_2<\rms$.  In other words, if $U(r)$ has one
root outside the outer horizon, then the other two roots are real and lie
between 0 and $\rms$.

\bigskip

\noindent
{\bf (2) If $U(\rps)>0$ then $U(r)>0$ for all $\rps\le r <\infty$}:
\medskip
 
  We have
\bea
U(\rps)= \fft{2(\rps-\rms)\, \rps}{p}\, \big[ p\, \rms + 
            (4n\, p -2p -16n)\, \rps\big]\,,
\eea
and hence if this is to be positive we must have
\bea
p=\fft{u\,\rms + 16 n\, \rps}{2(2n-1)\, \rps +\rms}\,,
\eea
with $u>0$.  Writing $\rps=\rms\, (1+w)$, it follows
that
\bea
U'(\rps) &=& \fft{\rms^3}{[2(2n-1)\rps + \rms]}\, \Big[
 48n + (8n+1)\, u + 2[64n^2+72n+(16n-1)\, u]\, w \nn\\
&& \qquad\qquad\qquad\qquad +
  8n\,(32n+28+3u)\, w^2 + 128n\,(n+1)\, w^3\Big]\,,
\eea
which is manifestly positive since $n\ge2$.
Since, as we showed above, $U(r)$ can have at most one zero for $r>\rps$,
and since $U(r)\longrightarrow +\infty$ as $r\longrightarrow +\infty$,
it follows that if $U(\rps)>0$ then $U(r)$ cannot be negative outside the
outer horizon.

  The conclusion from the properties established above is that if the
potential $V^-(r)$ is to be negative in some region outside the outer horizon,
then it must be that $U(r)$ would have to be negative on the horizon, i.e 
$U(\rps)<0$.  

   We have 
\bea
U(\rps) = 2(\rps-\rms)\, \rps\, W(\rps)\,,\qquad
  W(\rps) = \fft{p\, \rms - 2(p+8n-2n\, p)\,\rps}{p}\,.\label{Wdef}
\eea
If $U(\rps)<0$ it therefore follows that $W(\rps)<0$.  

\subsection{Lower bounds on $b^2$ for having $V^-(r)<0$ outside $\rps$}

    We know that $V^-(r)$ is always non-negative outside the outer horizon
if $b^2=1$, and we know from numerical examples that $V^-(r)$ can become
negative in some region outside the horizon if $b^2$ is sufficiently
large.  It is therefore of interest to determine bounds on $b^2$, and,
in particular, to determine the lowest value of $b^2$ for which the potential
can be negative outside the horizon.  This can be done as follows:

   We already showed that if $U(\rps)>0$ then $U'(\rps)>0$ and furthermore
$U(r)$ is then greater than zero outside the horizon.  Thus, in order to be
able to have $U(r)<0$ in some region outside the horizon, 
$U(\rps)$ would have to be negative.  The threshold case is then
when $U(\rps)=0$.  Solving this for $p$, we have that
\bea
U(\rps)=0 \quad \implies \quad p= \fft{16n\, \rps}{2(2n-1)\, \rps + \rms}\,.
\label{pthresh}
\eea
Substituting this value for $p$ into the cubic equation $\tilde\Lambda(p)$
given in eqn (\ref{pcubic}), we can solve for $b^2$, finding
\bea
b^2= \fft{[\rms + 2(2n+3)\,\rps]\, [3\rms^2 - 2\rms\,\rps +
    (n+1)(2n-1)\, \rps^2]}{16(n+1)\, \rms\,\rps\,
   [\rms + 2(2n-1)\,\rps]}\,.\label{solbsq}
\eea
It can easily be seen by writing 
\bea
\rps=\rms\, (1+w)\,,\qquad n=2+y\,,
\eea
that $b^2$ is positive for all valid values of the parameters (that is,
$y\ge0$ and $w\ge 0$).

   It can also easily be seen that viewing $b^2$ in eqn (\ref{solbsq})
as a function of $y$ and $w$, and differentiating with respect to $w$,
gives
\bea
\fft{\del b^2}{\del w} >0\,, \quad \forall\, w\ge 0\,.
\eea
This implies that the threshold value of $b^2$, above which the potential
can become negative outside the horizon, is as small as possible when $w$ 
approaches zero, that is, in the extremal limit
\bea
\rps \longrightarrow \rms\,.\label{extremal}
\eea
Setting $\rps=\rms$ in eqn (\ref{pthresh}) and eqn (\ref{solbsq}) then
implies that at threshold,
\bea
p= \fft{16n}{4n-1}\,,\qquad b^2 = \fft{(4n+7)(16n^2+8n-7)}{16(n+1)(4n-1)}\,.
\label{pbsq}
\eea
This gives the threshold values of $b^2$ for each $n=\ft12\ell(\ell+1)-1$.  
Thus we have for $\ell=2,3,4,5\cdots$ that 
\bea
b_{\rm min}=\{1.80525,2.53171,3.23273, 3.93299,\cdots\}\,.
\eea
If $b$ exceeds these values, then the potential will be negative
in some region outside the outer horizon.  Thus the smallest lower bound is
for the $\ell=2$ angular mode, and is $b> \sqrt{\dfft{365}{112}}$.

\section{Summary and Discussion}

  The goal of this paper was to study the linearised perturbations of
black hole solutions in theories of the kind that one encounters in 
supergravity.  Such theories typically involve scalar fields in addition
to the metric and Maxwell fields, and consequently the analysis of the
linearised perturbations is considerably more complicated than in 
pure Einstein or Einstein-Maxwell gravity.  Our focus in this paper has been
on the perturbations of static charged black holes in a three-parameter class
of Einstein-Maxwell-Dilaton-Axion (EMDA) theories, described by
the Lagrangian (\ref{emdalag}).

   A number of general features emerged in the course of our analysis
of the linearised perturbations.  The equations describing the dynamics
of the perturbed fields can be separated for all choices of the three
parameters in the EMDA Lagrangian.  That is to say, the angular and time
dependence can be factored off, thereby reducing the problem to the
analysis of ordinary differential equations satisfied by the radial 
functions.  The equations describing the perturbations in the
polar sector, which are disjoint from those for the axial sector, are
identical to those arising in the study of linearised perturbations in
a family of Einstein-Maxwell-Dilaton theories.  We studied these previously
in \cite{porowh}.  The polar perturbations were reduced to three functions
obeying three decoupled second-order radial equations.  
In the axial sector, we showed in the first five sections 
of the current paper that the perturbations are described by a coupled
system of three second-order radial equations, which cannot in general be
diagonalised.  Diagonalisation is possible in the special case where
two of the three parameters in the EMDA Lagrangian (\ref{emdalag}) are
fixed, namely $a=1$ and $c=-2$.

   By this means, we arrived at a one-parameter family of EMDA theories
\bea
{\cal L}= \sqrt{-g}\, \Big[ R -\ft12 (\del\phi)^2 - e^{\phi}\, F^2
-\ft12 e^{-2\phi}\,(\del\chi)^2 + 
   b\,\chi\,\widetilde{F}^{\mu\nu}F_{\mu\nu}\Big]\,,
\label{emdalagspec}
\eea
for which the linearised perturbations can be essentially fully analysed,
in the sense that they can be reduced to three axial functions obeying 
three decoupled second-order equations, and three polar functions obeying
three decoupled radial equations.  The remaining free parameter in the
family of EMDA theories (\ref{emdalagspec}) 
is the constant $b$ in the Lagrangian 
(\ref{emdalag}) (with $a=1$ and $c=-2$),
which characterises the strength of the axionic coupling to the 
topological term 
$\epsilon^{\mu\nu\rho\sigma}\, F_{\mu\nu}\, F_{\rho\sigma}$.

   We showed that the potentials $V^-_{(p)}(r)$ in the diagonalised radial
equations describing the linearised perturbations in the axial sector
are all non-negative outside the horizon when $b^2=1$, and in fact
for all values of $b^2$ that are sufficiently small.  If the potentials
are all non-negative, then this establishes the linearised mode stability
of the black hole solution.

  If, however, 
$b^2$ is larger than certain threshold values (which themselves
depend on the angular mode number $\ell$), then one of the three
axial potentials will become negative in a region outside, and near to, 
the outer horizon. Under such circumstances there no longer exists
a simple energy argument establishing mode stability, and a more
detailed investigation, perhaps numerical, would be needed in order
to see whether mode instabilities could then actually arise.  The threshold
values of $b^2$ are given in eqn (\ref{pbsq}).  Thus, taking $b$ to
be positive without loss of generality, then the smallest of the thresholds,
which occurs for the $\ell=2$ mode, is at $b=\sqrt{\dfft{365}{112}}
\approx 1.80525$.  If $b$ exceeds this value then one of the $\ell=2$ modes
in the axial sector has a potential that is negative  in a region
outside and near to the outer horizon.

  One specific member of the one-parameter family of theories 
described by the Lagrangian (\ref{emdalagspec}) is particularly 
notable, namely when $b=1$ (or, equivalently, and related by an orientation
reversal, $b=-1$).   We showed that in this specific case the
potentials $V_i^+(r)$  appearing in the three decoupled polar radial
equations $(\del_{r_*}^2 + \omega^2 - V^+_i)\, Z^+_i=0$ (for $i=1$, 2 and 3) 
are related to
the three potentials $V_i^-(r)$ appearing in the three decoupled axial radial
equations $(\del_{r_*}^2 + \omega^2 - V^-_i)\, Z^-_i=0$.  
In fact, for each value of $i$ the corresponding polar and axial potentials
are expressible in terms of a ``superpotential,''
with
\bea
V^\pm = \pm \del_{r_*} W + W^2 - \beta^2\,,
\eea
where $W=f+\beta$ and the superpotential function $f$ and constant 
$\beta$ are given in eqns (\ref{fsol}) and (\ref{albe}).  The axial and polar
fields are also related, with $D_\pm\, Z^\mp = (\im \omega \pm\beta)\,
Z^\pm$, where $D_\pm =\del_{r_*} \pm W$, as discussed earlier.

  Systems where there exists such a relation between two potentials are
known in the literature as ``supersymmetric quantum mechanical'' models.
(See, for example, \cite{nicolai1,nicolai2,witten1981,hayrau,valmorber,tong}.) 
Defining a supercharge 
\bea
\cQ= (\bp + \im W)\, \bb\,,
\eea
where $\bp=-\im \del_{r_*}$ is a momentum operator and $\bb$ is a fermionic
operator obeying
\bea
\{\bb^\dagger, \bb\} =1\,,\qquad \{\bb,\bb\}=0\,,\qquad
\{\bb^\dagger,\bb^\dagger\}=0\,,\qquad
[\bp,\bb]=0\,,\qquad [\bp,\bb^\dagger]=0\,,
\eea
it can be seen that $\{\cQ^\dagger,\cQ\}=2H$, with the Hamiltonian $H$ 
being given by
\bea
H=\ft12 (-\del_{r_*}^2+ V^+ +\beta^2)\bb\bb^\dagger +
  \ft12 (-\del_{r_*}^2+ V^- +\beta^2)\bb^\dagger \bb\,.
\eea
Representing $\bb$ by the matrix $\bb=\begin{pmatrix} 
 0&1\cr 0 &0\end{pmatrix}$, this implies
\bea
H= \fft12 \begin{pmatrix} -\del_{r_*}^2+ V^+ +\beta^2 & 0\cr
                  0& -\del_{r_*}^2+ V^- +\beta^2\end{pmatrix}\,,
\eea
and so by defining 
\bea
\bZ =\begin{pmatrix} Z^+ \cr Z^-\end{pmatrix}\,,
\eea
the polar and axial equations can be written in the unified form
\bea
H\, \bZ = \ft12 (\omega^2+\beta^2)\, \bZ\,.
\eea
  
   There is another special property of the one-parameter family of
Lagrangians (\ref{emdalagspec}) when $b=1$, namely that 
in this case it is a consistent truncation of
the bosonic theory given in eqn (\ref{truncSTUlag}).  This latter
theory is the bosonic sector of an ${\cal N}=2$ supergravity, which is
itself the consistent truncation of a larger ${\cal N}=2$ theory known
as STU supergravity.\footnote{It should be noted that although the
Maxwell field $H_{\mu\nu}$ in the bosonic Lagrangian (\ref{truncSTUlag})
can be consistently truncated by setting $H_{\mu\nu}=0$, thereby
yielding the EMDA theory described by the Lagrangian (\ref{emdalagspec}),
there is no analogous {\it supersymmetric} truncation of the
supergravity theory whose bosonic sector is given by (\ref{truncSTUlag}). 
That is to say, the bosonic theory described by the Lagrangian 
(\ref{emdalagspec}) is not {\it itself} the bosonic sector of
any supersymmetric theory.}  This supersymmetric consistent truncation of
STU supergravity is implemented in the bosonic sector 
by setting its four Maxwell fields equal
in pairs, yielding $F_{\mu\nu}$ and $H_{\mu\nu}$ in eqn (\ref{truncSTUlag}),
while at the same time setting two of its three dilatons to zero
and two of its three axions to zero.  (Corresponding truncations in the
fermionic sector must also be implemented.) 

  As we showed in section \ref{Hperturbations}, the axial and polar
radial functions $Z^\pm_{\sst H}$ associated with the perturbations of
the additional $H_{\mu\nu}$ field in the ${\cal N}=2$ supergravity theory
also have potentials $V^\pm_{\sst H}$ that are related via a superpotential.
Thus the feature of the $a=1$, $c=-2$, $b=1$ EMDA theory, of 
having axial and polar sectors that are related as a supersymmetric
quantum mechanical model, persists also for all the bosonic 
perturbations within the full ${\cal N}=2$ supergravity theory.

\section*{Acknowledgments}

We thank Dean Baskin, Mihalis Dafermos and Hermann Nicolai for helpful
discussions. We are grateful for hospitality at the Spring Workshop on
Gravity, Strings and Cosmology, Cook's Branch Conservancy,
Montgomery Texas, supported by the Mitchell
Family Foundation.  C.N.P and B.F.W. also gratefully acknowledge hospitality
and support from the Albert Einstein Institute, Potsdam. 
C.N.P.~and D.O.R.~are supported in part by DOE grant DE-SC0010813.  
B.F.W.~is supported in part by NSF grant PHY 1607323.

\appendix

\section{The Coupled $a=1\,, c=-2$
Second-Order Equations}

In section \ref{secondordersec} we obtained the system of three
coupled second-order equations describing the axial perturbations in
the 3-parameter EMDA theories in eqns (\ref{secondorderAxialeqs}), 
but we were unable to diagonalise the
equations for general values of the parameters $a$, $b$ and $c$.  We then
went on to study the system in terms of a set of six first-order equations,
and we showed how we could diagonalise the system if the special 
choice of parameters $a=1$ and $c=-2$ was made.  It is of interest now
to return to the direct second-order analysis of section \ref{secondordersec},
but now imposing the specialisation
\bea
a=1\,,\qquad c=-2\,.
\eea
With this done, we find that the coupled second-order system
given by eqns (\ref{secondorderAxialeqs}) and (\ref{tspcoefficients}) takes the
form
\bea
(\del_{r_*}^2 + \omega^2 + T_1)\, \Psi  + \fft{e^{2\bnu +\bar\phi}}{r^3}
\, {\bf M}\,\Psi=0\,,\qquad\qquad  
\Psi=\begin{pmatrix} \cG_{12}\cr B_{\u0\u3}\cr \chi\end{pmatrix}\,,
\label{3coupled}
\eea
where $T_1$ is the given in eqns (\ref{tspcoefficients}), but
now with $a=1$, and ${\bf M}$ is the constant matrix
\bea
{\bf M}= \begin{pmatrix} \rms + 3\rps & -2 \sqrt{2n}\, Q& 0\cr
         -2\sqrt{2n}\, Q & 0 & -2\sqrt{2(n+1)}\, b\, Q \cr
        0 & -2\sqrt{2(n+1)}\, b\, Q & \rms -\rps \end{pmatrix}\,.
\eea
Since the non-diagonal term in eqn (\ref{3coupled}) 
is of the form of an overall
function times the symmetric 
constant matrix ${\bf M}$, it is evident that the system can be
diagonalised by means of an orthogonal transformation.  The final
conclusion for the form of the resulting decoupled equations will be equivalent
to the decoupled equations that we obtained in section \ref{diagsec}.

   To make this more precise, the orthogonal matrix $S$ that 
diagonalises ${\bf M}$,
that is, ${\bf M}\longrightarrow S {\bf M} S^T={\bf M}_{\rm diag}$, 
has components $S_{ij}$ such that
\bea
\fft{S_{i1}}{S_{i2}} = \fft{ 2\sqrt{n}\, \sqrt{\rms\,\rps}}{(\rms + 3\rps -s_i)}\,,\qquad
\fft{S_{i3}}{S_{i2}} = -\fft{4n\, \rms\,\rps + (\rms + 3\rps -s_i)\, s_i}{
            2b\,\sqrt{n+1}\,\sqrt{\rms\,\rps}\,
      (\rms+3\rps -s_i)}\,,
\eea
for $i=1$, $2$ and $3$, 
where $s_i$ denotes the roots of the cubic equation
\bea
&&s^3 - 2 (\rms+\rps)\, s^2 + \Big(\rms^2 -3\rps^2 + 2\rms\,\rps\,
  \big[1-2n - 2(n+1)\, b^2\big]\Big)\, s \nn\\
&&+
  4\rms\, \rps\, \big[n(\rms-\rps) + (n+1)\, b^2\, (\rms+3\rps)\big] =0\,.
\label{scubic}
\eea
Comparing with the discussion in section \ref{diagsec}, it is evident that
we must have 
\bea
\fft{S_{i1}}{S_{i2}} = \fft{c_1}{c_2}\,,\qquad 
\fft{S_{i3}}{S_{i2}} = \fft{c_3}{c_2}\,,
\eea
where $c_1$ and $c_3$ are given in terms of $c_2$ in eqns 
(\ref{cm1top}) and (\ref{cm3top}), with $p$ taken to be the root $p_i$
of the cubic (\ref{pcubic}).  Indeed, it can be verified that these equations
hold, provided that $p$ is related to $s$ by 
\bea
s= \rms + \Big(3+\fft{8n}{p}\Big)\, \rps\,.\label{sprel}
\eea
In particular one finds that the cubic equation (\ref{scubic}) becomes
equivalent to the cubic equation (\ref{pcubic}) if eqn (\ref{sprel}) holds.

\section{\label{Plots} Some Example Potentials}

  Here we present some plots of the three axial
 potentials $V^-_i$ and the three polar potentials $V^+_i$ appearing in the
radial wave equations describing the linearised perturbations of the
static charged Gibbons-Maeda black holes, viewed as solutions in the
family of 
$a=1$, $c=-2$ EMDA theories described by the Lagrangian (\ref{emdalagspec}).
The six plots here show the potentials for $b=0$, $0.5$, $1$, $2$, $3$ and $4$,
as described in the caption to Fig.~1.  Note that the potentials all vanish
on the outer horizon, being proportional to $(r-\rps)$.  At large
$r$ the potentials all go to zero, with the asymptotic behaviour
$V\sim \dfft{\ell(\ell+1)}{r^2}$.

\begin{figure}
\ \ \ \ \ \includegraphics[width=6.5cm]{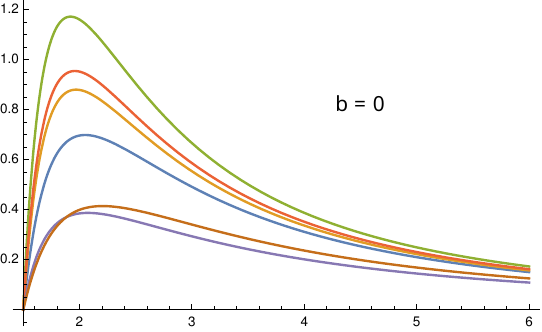}\ \ \ \ \ \ \ \ \
\includegraphics[width=6.5cm]{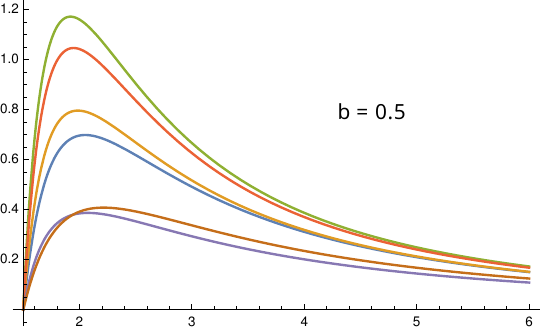}\\
\\ \\ \\ 
\phantom{M}  \ \ \includegraphics[width=6.5cm]{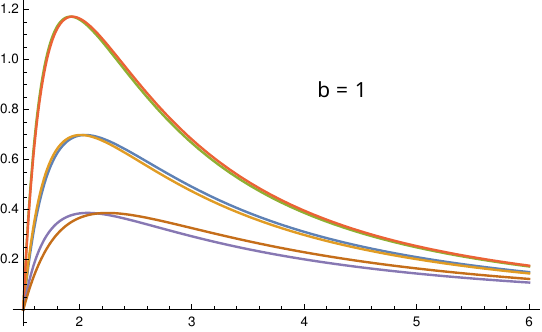}\ \ \ \ \ \ \ \ \
\includegraphics[width=6.5cm]{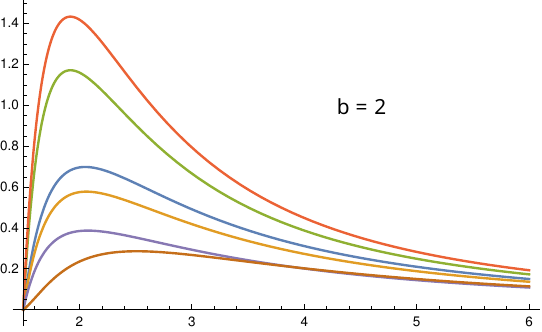}
\\ \\ \\
\phantom{M}  \ \ \includegraphics[width=6.5cm]{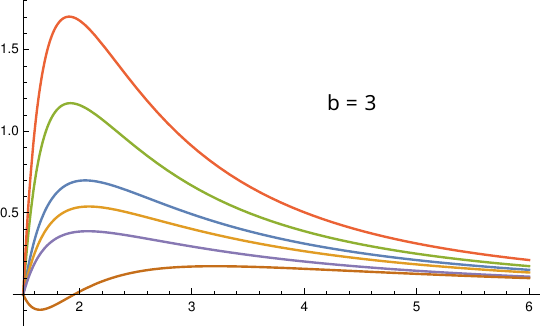}\ \ \ \ \ \ \ \ \
\includegraphics[width=6.5cm]{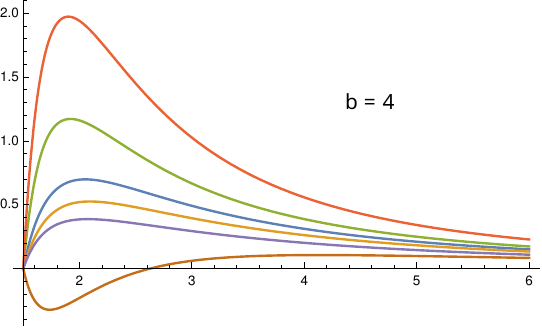}
\caption{\it Some example plots of the axial and polar potentials $
(V_1^-,V_1^+,V_2^-,V_2^+,V_3^-,V_3^+)$ corresponding to the roots
$(p_1,p_2,p_3)$ ordered as $p_1<p_2<p_3$, for
$\rps= 1.5\, \rms$, $\ell=2$, for various values of $b$, as indicated.
The dimensionless 
ratio $r/\rms$ is plotted from the outer horizon at $r/\rms=1.5$
to $r/\rms=6$ along the horizontal axis.  The dimensionless quantity
$V(r)\, \rms^2$ is plotted along the vertical axis, for the 
various potentials $(V_1^-,V_1^+,V_2^-,V_2^+,V_3^-,V_3^+)$.
The top pair
of almost superimposed potentials in the $b=1$ plots 
correspond to the middle root $p_2$, 
with the red line being the axial potential $V_2^-$.  The middle
pair of almost superimposed $b=1$ plots correspond to the the smallest 
root $p_1$, with the orange line being the axial potential $V_1^-$.
The bottom pair of almost superimposed $b=1$ plots correspond to the 
largest root $p_3$ (that is, the positive root), with the brown line being
the axial potential $V_3^-$.  One can see how the potentials change as
$b$ is adjusted, and in particular how
the axial potential $V_3^-$ becomes negative
in a region outside the outer horizon when $b$ is large enough.}
\label{Fig1-6}
\end{figure}

   In the case that $b=1$ we can also include the potentials 
$V^\pm_{\sst H}$ for the axial and polar perturbations of the additional
Maxwell field $H_{\mu\nu}$ of the ${\cal N}=2$ supergravity theory
whose bosonic sector is described by the Lagrangian (\ref{truncSTUlag}). 
This gives the plots shown in Fig.~2.

\begin{figure}
\centerline{\includegraphics[width=8.5cm]{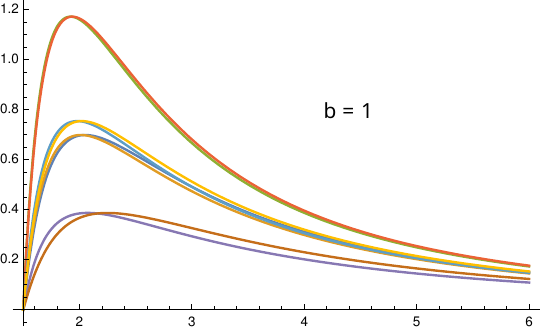} }
\caption{\it This is the same as the middle-left panel above, showing the
potentials $(V_1^-,V_1^+,V_2^-,V_2^+,V_3^-,V_3^+)$ for $b=1$, but now with
the potentials $(V^-_{\sst H}, V^+_{\sst H})$ included as well.  These 
are the pair of lines that are just above the previous middle pair
$(V_1^-,V_1^+)$.  The light blue line is the axial potential $V^-_{\sst H}$.}
\label{Fig7}
\end{figure}

\end{document}